\documentclass[draftclsnofoot,onecolumn,12pt]{IEEEtran}
\IEEEoverridecommandlockouts
\usepackage{amsmath,amssymb,amsfonts}
\usepackage{algorithmic}
\usepackage{amsmath} 
\usepackage{graphicx}
\usepackage{textcomp}
\usepackage{xcolor}
\usepackage{amsmath}
\renewcommand{\vec}[1]{\mathbf{#1}}
\usepackage{algorithm}
\usepackage{algorithmic}
\usepackage{graphicx}
\usepackage{xcolor}
\usepackage{float} 
\usepackage{cite}
\usepackage{stfloats}
\usepackage[section]{placeins}
\usepackage{subfigure}
\usepackage{booktabs}
\usepackage[numbers,sort&compress]{natbib}
\usepackage{amssymb}
\usepackage{setspace}
\usepackage{multirow}
\usepackage{color}
\usepackage{hyperref}
\usepackage{pifont}
\usepackage{bm}

\newtheorem{remark}{Remark}

\usepackage{titlesec}

\DeclareMathOperator{\Tr}{Tr}
\fontsize{5.0pt}{\baselineskip}\selectfont
\usepackage{epsfig}
\def\BibTeX{{\rm B\kern-.05em{\sc i\kern-.025em b}\kern-.08em
    T\kern-.1667em\lower.7ex\hbox{E}\kern-.125emX}}

\begin{document}
% \doublespacing
\addtolength{\topmargin}{0.04in}
% \title{Maximum Likelihood Algorithm for ST-CDMA LFMCW MIMO radar Systems
% }
\title{CRB Analysis for Mixed-ADC Based DOA Estimation
}
\author{Xinnan Zhang, Yuanbo Cheng, Xiaolei Shang, and Jun Liu, \emph{Senior Member, IEEE}% <-this % stops a space
% \thanks{This work was supported in part by the  National Natural Science Foundation of China under Grant 61771442, in part by Key Research Program of Frontier Sciences of CAS under Grant QYZDY-SSW-JSC035, and in part by the Swedish Research Council (VR grants 2017-04610  and 2016-06079).}% <-this % stops a space
\thanks{X. Zhang, Y. Cheng, X. Shang, and J. Liu are  with the Department of Electronic Engineering and Information
Science, University of Science and Technology of China, Hefei 230027, China
(e-mail: zhangxinnan@mail.ustc.edu.cn; cyb967@mail.ustc.edu.cn; xlshang@mail.ustc.edu.cn; junliu@ustc.edu.cn).}
\thanks{Portions of this work were presented at the 2023 IEEE International Conference on Acoustics, Speech, and Signal Processing \cite{zhang2023ula}.}}

% \thanks{J. Li is with the Department of Electrical and Computer Engineering, University of Florida, Gainesville, FL 32611 USA (e-mail: j.li@ieee.org). }
% \thanks{P. Stoica is with the Department of Information Technology, Uppsala University, Uppsala SE-751 05, Sweden (e-mail: ps@it.uu.se).} 
% }
\maketitle
\begin{abstract}

We consider a mixed analog-to-digital converter (ADC) based architecture consisting of high-precision and one-bit ADCs \textcolor{black}{with the antenna-varying threshold for direction of arrival (DOA) estimation using a uniform linear array (ULA), which utilizes fixed but different thresholds for one-bit ADCs across different receive antennas.} The Cram{\'e}r-Rao bound (CRB) with the antenna-varying threshold is obtained. Then based on the lower bound of the CRB, we derive the asymptotic CRB of the DOA, which depends on the placement of mixed-ADC.
Our analysis shows that distributing high-precision ADCs evenly around the two edges of the ULA yields improved performance. This result can be extended to a more general case where the ULA is equipped with two types of ADCs with different quantization precisions. To efficiently obtain the maximum likelihood DOA estimates, we propose a two-step algorithm. Firstly, we formulate the model as a sparse signal representation problem, and modify the sparse learning via iterative minimization (SLIM) approach to the mixed-ADC based DOA estimation. In the second step, we use the \textcolor{black}{relaxation-based approach} to cyclically refine the estimates of SLIM, further enhancing the DOA estimation performance. Numerical examples are presented to demonstrate the validity of the CRB analysis and the effectiveness of our methods.
\end{abstract}
\section{Introduction}
Direction of arrival (DOA) estimation  is a fundamental problem in array signal processing, and has wide-ranging applications in sonar, radar, navigation, and more \cite{tuncer2009classical,sun2020mimo}. In particular, DOA estimation is commonly employed in radar systems to detect and identify objects such as vehicles, pedestrians, and obstacles on the road by providing angle information, which is crucial for advanced driver assistance systems and autonomous driving.

Over the past few decades,  the field of DOA estimation has received substantial attention from researchers, leading to the development of numerous methods.  Traditional subspace-based methods, e.g., the multiple signal classification \cite{schmidt1982signal} and the estimation of parameters by rotational invariant techniques \cite{roy1989esprit}, require prior knowledge of the number of sources and a significant number of snapshots to achieve satisfactory performance. \textcolor{black}{The advancements in utilizing sparse representations of signals have led to the creation of various compressed-sensing algorithms \cite{yang2018sparse}, with on-grid algorithms standing out for their assumption that DOAs are on predefined grids. The compressed sensing algorithms often involve specific hyperparameters, which are hard to choose in practice. The non-parametric IAA \cite{5417172}, along with semi-parametric adaptive algorithms like SPICE \cite{stoica2010spice,stoica2014weighted}, LIKES \cite{stoica2012spice}, and SLIM \cite{tan2010sparse}, have been proven to be highly effective. These methods not only achieve super resolution but also avoid the issue of requiring hyperparameters. Furthermore, these algorithms can be used with Bayesian information criterion (BIC) to effectively estimate the number of signal sources and their parameters, regardless of whether the incoming signals are correlated or coherent, or whether the number of shapshots is one or more. These algorithms rely on data sampled by high-precision analog-to-digital converters (ADCs), showing promise in improving DOA estimation accuracy while reducing the required snapshot numbers.}
% The development in the technology of sparse representation has enabled the proposal of a range of algorithms \cite{5417172,stoica2014weighted} that rely on data sampled by high-precision analog-to-digital converters (ADCs), showing promise in improving DOA estimation accuracy while reducing the required snapshot numbers. 
Nevertheless, the power consumption and hardware cost of an ADC increase exponentially as the quantization bit and sampling rate grow  \cite{walden1999analog}, limiting their practical use of certain applications such as phase-modulated continuous-wave (PMCW) multiple-input multiple-output (MIMO) radar \cite{sun2020mimo,4350230,li2008mimo}. 

Using low resolution ADCs (e.g., 1-4 bits) \cite{singh2009limits,liu2017one,1039405,huang2019one,sedighi2021performance,shang2021weighted} has recently emerged as a promising technique to mitigate the aforementioned problem, and has been widely studied due to its low cost and low power consumption benefits. In particular, for a one-bit system, one-bit sampling with time-varying known thresholds to obtain signed measurements, has been considered in \cite{8335511,8822763,8291043}, but its implementation cost is high. Alternatively, one-bit sampling with antenna-varying thresholds \cite{liu2018massive, liu2020angular} has been proposed to reduce implementation cost while maintaining performance, by utilizing fixed \textcolor{black}{but different} thresholds for one-bit ADCs across different receive antennas. \textcolor{black}{It selects the thresholds once randomly and fixes all time for each antenna output.} Despite their potential advantages, low resolution ADC systems suffer from dynamic range problems, i.e., a strong target can mask a weak target \cite{walden1999analog}, which is especially crucial in advanced automotive applications.

\begin{figure}[htbp]
\centering
\includegraphics[scale=0.095]{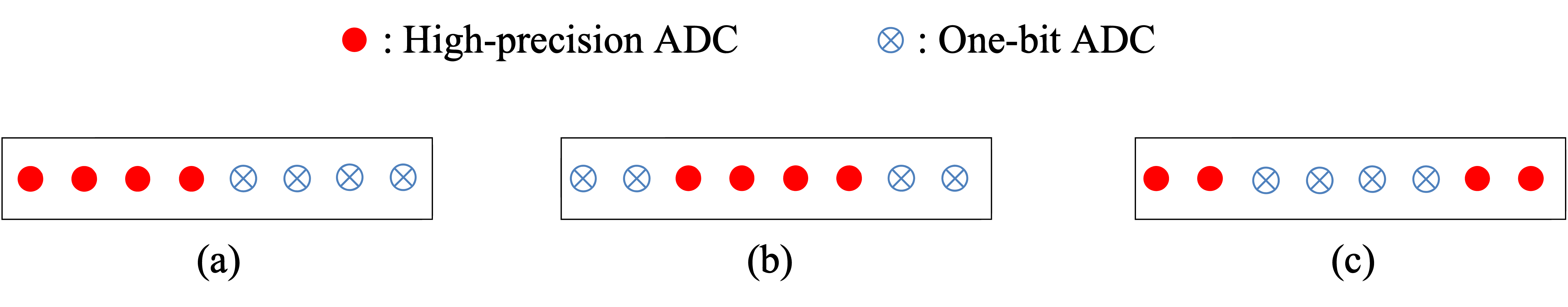}
\caption{Different placement in mixed-ADC model using ULA}
\label{fig:arr}
\end{figure}

A mixed-ADC based architecture was proposed in \cite{liang2016mixed} as a solution to overcome the limitations of the one-bit system, in which most receive antennas' outputs are sampled by one-bit ADCs and a few by high-resolution ADCs. Subsequently, \cite{shi2022performance} derived the closed-form expression for the Cram{\'e}r-Rao bound (CRB) and  demonstrated that the subspace-based methods can be utilized without any modification when using a uniform linear array (ULA).  In \cite{10027928,10414289}, the authors performed a CRB study in PMCW MIMO radar. To address the channel estimation problem, a gridless atomic norm based approach was proposed in \cite{wang2018gridless}. However, most of works make use of additive quantization noise model to approximate the distortion to noise caused by one-bit ADCs, or simply employ high-precision ADCs  and one-bit ADCs on different sides of the ULA (see the first case in Fig. \ref{fig:arr} for an example), without considering their relative placement. Although the impacts of the placement and the number of high-precision ADCs were investigated through the MSEs under four configurations on channel estimation problems \cite{wang2018gridless}, the study lacks theoretical justifications.

In this study, we focus on the theoretical analysis of the placement of high-precision and one-bit ADCs when using the ULA for DOA estimation. We use the deterministic model \cite{60109}, where the signal matrix is assumed to be deterministic but unknown, to analyze the CRB of the DOA estimation performance. Our main contributions can be summarized as follows:
\begin{enumerate}
    \item We first derive the CRB under the mixed-ADC architecture. Based on the lower bound of the CRB, we further derive the asymptotic CRB of the DOA and discover that it is related to the placement of high-precision and one-bit ADCs, which has not been previously explored.  Then we introduce the placement of mixed-ADC problem, and prove that the high-precision ADC pairs (each antenna output needs a pair of ADCs for the I and Q channels) should be distributed evenly around the edges of ULA to achieve the best performance when using ULAs.
    
    \item The impact of the proportion of high-precision ADC’s pairs with the optimal placement is investigated using a performance efficiency factor we introduce under the mixed-ADC architecture.

    \item A maximum likelihood (ML) method is adopted to estimate DOAs. Instead of optimizing ML directly, we modify the sparse parameters via iterative minimization (SLIM) to the mixed-ADC data model. In addition, we use the majorization-minimization (MM) technique to iteratively update the parameters in an efficient manner.
    
    \item We design a two-step approach, referred to as SLIM-RELAX, to further improve the performance of SLIM and extend the BIC to mixed-ADC model to determine the number of signals.
\end{enumerate}

The rest of the paper is organized as follows. Section \ref{sec:2} gives the signal model under mixed-ADC architecture. The CRB of the DOA for mixed-ADC architecture is derived and analyzed in Section \ref{sec:3}. In Section \ref{sec:6}, we introduce a two-step algorithm for accurate DOA estimation. Numerical examples are presented in Section \ref{sec:7}.

\indent \textit{Notation:} We denote vectors and matrices by bold lowercase and uppercase letters, respectively.  $(\cdot)^T$ and $(\cdot)^H$ represent the  transpose and the conjugate transpose, respectively. $\vec{R}\in \mathbb{R}^{N\times M}$ or  $\vec{R}\in \mathbb{C}^{N\times M}$ denotes a real or complex-valued $N\times M$ matrix $\vec{R}$. $\vec{I}_N$ is an $N\times N$ identity matrix and $\mathbf{1}_N=[1, \ldots, 1]^T \in \mathbb{R}^{N \times 1}$.
$\vec{A}\otimes\vec{B}$, $\vec{A}\odot\vec{B}$ and $\vec{A} \ast \vec{B}$ denote the Kronecker, Hadamard and Khatri–Rao matrix products, respectively. $\mathbf{A}[\bm{\delta},:]$ extracts rows and elements corresponding to the non-zero values of the logical array $\bm{\delta}$. $\|\cdot\|_2$ denotes the $\ell_2$ norm, $\rm{vec}(\cdot)$  refers to the column-wise vectorization
operation, and $\rm{diag}(\vec{d})$ is a diagonal matrix with diagonal
entries formed from $\vec{d}$. 
$A_{\mathrm{R}}\triangleq \Re\{A\}$ and
$A_{\mathrm{I}} \triangleq \Im\{A\}$, where $\Re\left\{ \cdot\right\}$ and $\Im\left\{ \cdot\right\}$ denote the real and imaginary parts, respectively. $\mathrm{sign}(\cdot)$ is the sign function applied element-wise to vector or matrix, and $\lfloor \cdot \rfloor$ is the floor function. Finally, $\mathrm{i} \triangleq \sqrt{-1}$.

\section{Signal model}\label{sec:2}
We consider $K$ narrowband far-field signals impinging on a ULA with $M$ elements from directions $\{\theta_1, \dots, \theta_K\} $. After being sampled and quantized with ADCs, the array output vector $\mathbf{x}(n)=[\text{x}_1(n), ..., \text{x}_M(n)]^T\in \mathbb{C}^{M\times1}$ can be expressed as
\begin{align}
    \mathbf{x}(n)=\mathbf{A} \mathbf{s}(n)+\mathbf{e}(n),
\end{align}
where $n$ is the snapshot index, $n=1,2,\dots,N$, with $N$ being the total number of snapshots, $\mathbf{s}(n) = [\text{s}_1(n), ..., \text{s}_K(n)]^T \in \mathbb{C}^{K\times 1}$ is the signal vector, and $\mathbf{A}=\left[\mathbf{a}\left(\theta_1\right), \cdots, \mathbf{a}\left(\theta_K\right)\right] \in \mathbb{C}^{M \times K}$ represents the array steering matrix with $\mathbf{a}(\theta_k)$ denoting the steering vector associated with the $k$th source. The noise vector $\mathbf{e}(n) = [\text{e}(n),...,\text{e}_M(n)]\in \mathbb{C}^{M\times 1}$ is the zero-mean circularly symmetric complex-valued white Gaussian noise with independent and identically distributed (i.i.d.) $\mathcal{C N}\left(0, \sigma^2\right)$ entries. By letting $\omega_k = \pi\sin\theta_k$, the steering vector $\mathbf{a}(\omega_k)$ can be written as
\small
\begin{align}
    \vec{a}(\omega_k)&=\begin{bmatrix} 1, e^{\mathrm{i}2\frac{d}{\lambda}\omega_k},  \dots, e^{\mathrm{i}2\frac{d}{\lambda}(M-1)\omega_k}\end{bmatrix}^T,  \quad k=1,2,...,K,
\end{align}
\normalsize
where $d$ is the inter-element spacing of the ULA and $\lambda$ is the signal wavelength. 

Let $\mathbf{X} = [\mathbf{x}(1), \mathbf{x}(2), \ldots, \mathbf{x}(N)] \in \mathbb{C}^{M \times N}$ represent the received signal matrix. We can represent $\mathbf{X}$ as 
\begin{align}\label{eq:3}
    \mathbf{X} = \mathbf{A}\mathbf{S}+\mathbf{E},
\end{align}
where $\mathbf{S}=[\mathbf{s}(1), \mathbf{s}(2), \ldots, \mathbf{s}(N)] \in \mathbb{C}^{ K \times N}$ represents the signal matrix and $\mathbf{E}=[\mathbf{e}(1), \mathbf{e}(2), \ldots, \mathbf{e}(N)] \in \mathbb{C}^{M \times N}$ is the noise matrix. The signal matrix $\mathbf{S}$ is assumed to be deterministic unknown, making (\ref{eq:3}) a conditional or deterministic data model \cite{60109}.

When the antenna outputs are sampled by one-bit ADCs  with antenna-varying thresholds, we obtain
\begin{align}\label{eq:4}
    \mathbf{Z} = \mathcal{Q}(\mathbf{X} - \mathbf{H}),
\end{align}
where $\mathbf{H} \in \mathbb{C}^{M \times N}$ represents the known antenna-varying threshold, and $\mathcal{Q}(\cdot) = \mathrm{sign}(\Re\{\cdot\})+\mathrm{i} \cdot \mathrm{sign}(\Im\{\cdot\})$ denotes the complex one-bit quantization operator.

Consider now a mixed-ADC architecture  equipped with $M_0$ pairs of high-resolution ADCs and  $M_1$ pairs of one-bit ADCs. Here,  $M_0 + M_1 = M$, and we define $\kappa \triangleq M_0/M$  as the proportion of high-precision ADCs in the mixed-ADC architecture. We use the high-precision ADC indicator vector $\bm{\delta} = [\delta_1, \dots, \delta_M]^T$, where $\delta_i \in \{0, 1\}$ indicates that the $i$th antenna output is sampled by a pair of high-precision ADCs when $\delta_i = 1$ and  by one-bit ADCs when $\delta_i = 0$. The output of the mixed-ADC architecture can then be represented as
\begin{equation}\label{eq:5}
   \mathbf{Y} = \mathbf{Z} \odot (\boldsymbol{\bar{\delta}} \otimes \vec{1}_N^T) + \mathbf{X} \odot (\boldsymbol{\delta} \otimes \vec{1}_N^T),
\end{equation}
where $\boldsymbol{\bar{\delta}} = \mathbf{1}_M - \boldsymbol{\delta}$ is the indicator for the one-bit ADCs. \textcolor{black}{The output of the high-precision ADCs and the corresponding array steering vectors are denoted as $\mathbf{Y}_0 = \mathbf{Y}[\boldsymbol{\delta, :}] \in \mathcal{C}^{M_0 \times N}$ and $ \mathbf{A}_0 = \mathbf{A}[\boldsymbol{\delta, :}]$, respectively. Similarly, the one-bit counterparts are denoted as $\mathbf{Y}_1 = \mathbf{Y}[\boldsymbol{\bar{\delta}, :}] \in \mathcal{C}^{M_1 \times N}$ and $ \mathbf{A}_1 = \mathbf{A}[\boldsymbol{\bar{\delta}, :}]$.} Our problem of interest is to determine the DOA vector $\bm{\theta}=[\theta_1,...,\theta_K]^T$ from $\mathbf{Y}$.

\section{CRAMÉR-RAO BOUND ANALYSIS}\label{sec:3}

Let $\bm{\varphi}$ denote the vector of real-valued unknown target parameters, such that $\bm{\varphi}=\begin{bmatrix} \bm{\theta}^T,\ \vec{s}_{\rm R}^T, \ \vec{s}_{\rm I}^T \end{bmatrix}^T\in \mathbb{R}^{(K+2KN)\times 1}$, where $ \vec{s} = \rm{vec}(\mathbf{S})$. In the case where the noise variance   $\sigma^2$ is unknown,  the parameter vector becomes $\bm{\chi}=\begin{bmatrix}\bm{\varphi}^T, \ \sigma\end{bmatrix}^T\in \mathbb{R}^{(K+2KN+1)\times 1}$. Note that $\vec{E}$ is the circularly symmetric complex-valued white Gaussian noise with  i.i.d. $\mathcal{CN}(0,\sigma^2)$ entries. Thus the likelihood function of the measurement matrix $\vec{Y}$ can be expressed as
\begin{equation}\label{eq:6}
\ln{L}(\boldsymbol{\chi})= \ln{L}_0(\boldsymbol{\chi})+\ln{L}_1(\boldsymbol{\chi}),
\end{equation}
where $ L_0(\boldsymbol{\chi})$ and $ L_1(\boldsymbol{\chi})$ are the likelihood functions of high-precision data and one-bit data, respectively. The Fisher information matrix (FIM) for the mixed-ADC based data model is given by

\begin{align}\label{eq:7}
\mathbf{F}_{\mathrm{m}}=\mathbb{E}\left[\frac{\partial \ln L(\boldsymbol{\chi})}{\partial \boldsymbol{\chi}} \frac{\partial \ln L(\boldsymbol{\chi})}{\partial \boldsymbol{\chi}^T}\right].
\end{align}
By substituting (\ref{eq:6}) into (\ref{eq:7}), we can rewrite the FIM as
\small
\begin{align} \label{eq:8}
\mathbf{F}_{\mathrm{m}}&= \mathbb{E}\left[\frac{\partial \ln L_{0}(\boldsymbol{\chi})}{\partial \boldsymbol{\chi}} \frac{\partial \ln L_{0}(\boldsymbol{\chi})}{\partial \boldsymbol{\chi}^{T}}\right]+\mathbb{E}\left[\frac{\partial \ln L_{1}(\boldsymbol{\chi})}{\partial \boldsymbol{\chi}} \frac{\partial \ln L_{1}(\boldsymbol{\chi})}{\partial \boldsymbol{\chi}^{T}}\right]  \nonumber \\&+\mathbb{E}\left[\frac{\partial \ln L_{0}(\boldsymbol{\chi})}{\partial \boldsymbol{\chi}} \frac{\partial \ln L_{1}(\boldsymbol{\chi})}{\partial \boldsymbol{\chi}^{T}}\right]+\mathbb{E}\left[\frac{\partial \ln L_{1}(\boldsymbol{\chi})}{\partial \boldsymbol{\chi}} \frac{\partial \ln L_{0}(\boldsymbol{\chi})}{\partial \boldsymbol{\chi}^{T}}\right].
\end{align}
\normalsize

Note that the noise terms in the likelihood functions are independent of each other for high-precision and one-bit data. Therefore, we have:
\begin{align}
    &\mathbb{E}\left[\frac{\partial \ln L_{0}(\boldsymbol{\chi})}{\partial \boldsymbol{\chi}} \frac{\partial \ln L_{1}(\boldsymbol{\chi})}{\partial \boldsymbol{\chi}^{T}}\right] \nonumber\\&=\underbrace{\mathbb{E}\left[\frac{\partial \ln L_{0}(\boldsymbol{\chi})}{\partial \boldsymbol{\chi}}\right]}_{=\bm{0}} \underbrace{\mathbb{E}\left[\frac{\partial \ln L_{1}(\boldsymbol{\chi})}{\partial \boldsymbol{\chi}^{T}}\right]}_{=\bm{0}}.
\end{align}
This simplifies the expression for $\mathbf{F}_{\mathrm{m}}$ to
\small
\begin{align}\label{eq:10}
    \mathbf{F}_{\mathrm{m}}=\mathbb{E}\left[\frac{\partial \ln L_{0}(\boldsymbol{\chi})}{\partial \boldsymbol{\chi}} \frac{\partial \ln L_{0}(\boldsymbol{\chi})}{\partial \boldsymbol{\chi}^{T}}\right]+\mathbb{E}\left[\frac{\partial \ln L_{1}(\boldsymbol{\chi})}{\partial \boldsymbol{\chi}} \frac{\partial \ln L_{1}(\boldsymbol{\chi})}{\partial \boldsymbol{\chi}^{T}}\right].
\end{align}
\normalsize

Therefore, the FIM for the mixed-ADC data based model can be obtained by  summing the FIMs for the high-precision and one-bit data. Using these FIMs, we can obtain the CRB matrix for the mixed-ADC data model.

\subsection{Fisher Information Matrix for High-precision Data}
    \subsubsection{Known noise variance}
    The FIM for high-precision data is presented in \cite{stoica2005spectral}
    \begin{align}
        \mathbf{F}_{0}(\boldsymbol{\varphi})=\frac{2}{\sigma^{2}} \Re\left\{\mathbf{U U}^{H}\right\},
    \end{align}
    where 
    % \begin{align}
    %     &\mathbf{U}=\left[\boldsymbol\Delta, \quad\mathbf{G}, \quad \mathrm{i} \mathbf{G}\right ]^{H}, \\
    %     &\boldsymbol\Delta =\mathbf{S}^T \ast\dot{\mathbf{A}}, \quad \mathbf{G} = \mathbf{I}_N \otimes \mathbf{A}, \\
    %     &\dot{\mathbf{A}}=\left[\frac{\partial \mathbf{a}\left(\omega_1\right)}{\partial \omega_1}, \ldots, \frac{\partial \mathbf{a}\left(\omega_K\right)}{\partial \omega_K}\right]. 
    % \end{align}
    \begin{align}
        \mathbf{U}=\left[\boldsymbol\Delta, \quad\mathbf{G}, \quad \mathrm{i} \mathbf{G}\right ]^{H}
    \end{align}
    with 
    \begin{align}
        &\boldsymbol\Delta =\mathbf{S}^T \ast\dot{\mathbf{A}}, \quad \mathbf{G} = \mathbf{I}_N \otimes \mathbf{A}, \\
        &\dot{\mathbf{A}}=\left[\frac{\partial \mathbf{a}\left(\omega_1\right)}{\partial \omega_1}, \ldots, \frac{\partial \mathbf{a}\left(\omega_K\right)}{\partial \omega_K}\right]. 
    \end{align}
    \subsubsection{Unknown noise variance}
    In the case where the noise variance is unknown, the FIM for the  unknown parameter vector $\bm{\chi}$ is given by
    \begin{align}
        \mathbf{F}_0(\bm{\chi})=\frac{2}{\sigma^2}\left[\begin{array}{cc}
        \Re\{\mathbf{U}\mathbf{U}^H\} & \mathbf{0} \\
        \mathbf{0} & 2N M
        \end{array}\right].
    \end{align}

\subsection{Fisher Information Matrix for One-bit Data}

    \subsubsection{Known noise variance}
    Li {\textit{et al.} in \cite{li2018bayesian} established a connection between FIMs for high-precision data and one-bit data. The FIM for one-bit data is written as
    \begin{align}
        \mathbf{F}_1(\bm{\varphi}) = \frac{1}{\pi \sigma^{2}}(\mathbf{U}_{\rm R} \boldsymbol{\Lambda}_{\rm R} \mathbf{U}_{\rm R}^{T}+\mathbf{U}_{\rm I} \boldsymbol{\Lambda}_{\rm I} \mathbf{U}_{\rm I}^{T}),
    \end{align}
    where $\boldsymbol{\Lambda}=\text{diag}\left(\left[\lambda_{1}, \ldots, \lambda_{MN}\right]\right)$. The diagonal element $\lambda_k$ in $\boldsymbol{\Lambda}$ is given by 
    \begin{align}
        \lambda_{k}=B\left(\frac{\Re\left(\zeta_{k}\right)}{\sigma / \sqrt{2}}\right)+\mathrm{i} B\left(\frac{\Im\left(\zeta_{k}\right)}{\sigma / \sqrt{2}}\right),
    \end{align}
    where $\zeta_k$ is the $k$th element in  $\bm{\zeta}=\text{vec}(\mathbf{A}\mathbf{S}-\mathbf{H}) \in \mathbb{C}^{MN \times 1}$ and the function $B(\cdot)$ is defined by
    \begin{align}\label{eq:19}
        B(x)=\left[\frac{1}{\Phi(x)}+\frac{1}{\Phi(-x)}\right] e^{-x^{2}}
    \end{align}
    with $\Phi(x)=\int_{-\infty}^x \frac{1}{\sqrt{2 \pi}} e^{-\frac{t^2}{2}} d t$ being the cumulative distribution function of the standard normal distribution.
    
    \subsubsection{Unknown noise variance}
    For the case of unknown noise standard deviation $\sigma$, the matrix $\mathbf{U}$ is modified as
    \begin{align}\label{eq:20}
        \tilde{\mathbf{U}} = \left[\mathbf{U}^T; \bm{\zeta}/\sigma 
        \right]^T,
    \end{align}
    and the resulting FIM for the parameter vector $\bm{\chi}$ is given by
    \begin{align}
        \mathbf{F}_1(\bm{\chi}) = \frac{1}{\pi \sigma^{2}}(\tilde{\mathbf{U}}_{\rm R} \boldsymbol{\Lambda}_{\rm R} \tilde{\mathbf{U}}_{\rm R}^{T}+\tilde{\mathbf{U}}_{\rm I} \boldsymbol{\Lambda}_{\rm I} \tilde{\mathbf{U}}_{\rm I}^{T}).
    \end{align}
    
    \subsection{Fisher Information Matrix for Mixed Data}
    From the special structure of the matrix $\mathbf{U}$, we can obtain the following relations:
    \begin{align}
    \mathbf{U}_0 &=\mathbf{U} \text{diag}\left(\mathbf{1}_N \otimes \boldsymbol{\delta}\right), \nonumber\\
    \mathbf{U}_1 &=\mathbf{U} \text{diag}\left(\mathbf{1}_N \otimes \overline{\boldsymbol{\delta}}\right).
    \end{align}
    % respectively.
    \subsubsection{Known noise variance}
    
    Following the result in (\ref{eq:10}), we obtain the FIM for the mixed-ADC based data as
    \begin{align}
        \mathbf{F}_{\mathrm{m}}(\bm{\varphi}) = &\frac{2}{\sigma^{2}} \Re\left\{\mathbf{U}_{0} \mathbf{U}_{0}^{H}\right\} \nonumber \\ &+
        \frac{1}{\pi \sigma^{2}}\left(\mathbf{U}_{1, \rm R} \boldsymbol{\Lambda}_{\rm R} \mathbf{U}_{1, \rm R}^{T}+\mathbf{U}_{1, \rm I} \boldsymbol{\Lambda}_{\rm I} \mathbf{U}_{1, \rm I}^{T}\right),
    \end{align}
    where the first and second terms of $\mathbf{F}_{m}$ represent the FIMs of high-precision data and one-bit data, respectively.

    \subsubsection{Unknown noise variance}
    When the noise variance $\sigma^2$ is unknown, we have, likewise,
    \begin{align}
        \mathbf{F}_{\mathrm{m}}(\bm{\chi}) &= \frac{2}{\sigma^2}\Re\left\{\left[\begin{array}{cc}
        \mathbf{U}_0\mathbf{U}_0^H & \mathbf{0} \\
        \mathbf{0} & 2N M_0
        \end{array}\right]\right\}  \nonumber \\&+\frac{1}{\pi \sigma^{2}}\left(\tilde{\mathbf{U}}_{1, \rm R} \boldsymbol{\Lambda}_{1, \rm R} \tilde{\mathbf{U}}_{1, \rm R}^{T}+\tilde{\mathbf{U}}_{1, \rm I} \boldsymbol{\Lambda}_{1, \rm I} \tilde{\mathbf{U}}_{1, \rm I}^{T}\right).
    \end{align}
    Thus, the CRB of the mixed-ADC based data for the unknown parameter vector $\bm{\chi}$ can be computed by 
    \begin{align}\label{eq:24}
        \vec{CRB}(\bm{\chi}) = \mathbf{F}_{\mathrm{m}}^{-1}(\bm{\chi}).
    \end{align}

    \subsection{Asymptotic CRB for DOA}
    Based on the observation that the function $B(\cdot)$ is bounded by the interval $(0, 4]$ and reaches its maximum at the point 0 \cite{li2018bayesian}, we can obtain the following inequality:
    \begin{align}
        \mathbf{F}_{\mathrm{m}}(\bm{\chi}) \preceq &  \frac{2}{\sigma^2}\Re\left\{\left[\begin{array}{cc}
        \mathbf{U}_0\mathbf{U}_0^H & \mathbf{0} \\
        \mathbf{0} & 2N M_0
        \end{array}\right]\right\}  \nonumber \\&+\frac{4}{\pi \sigma^{2}}\left(\tilde{\mathbf{U}}_{1, \rm R} \tilde{\mathbf{U}}_{1, \rm R}^{T}+\tilde{\mathbf{U}}_{1, \rm I} \tilde{\mathbf{U}}_{1, \rm I}^{T}\right),
    \end{align}
    where the equality is achieved when the threshold $\mathbf{H} = \mathbf{AS}$. While this threshold is not feasible in practice, it  provides a theoretical upper bound for the FIM. Consequently, the lower bound of $\vec{CRB}(\bm{\varphi})$, denoted as $\vec{CRB}_\mathrm{L}(\bm{\varphi})$, can be expressed as
    \begin{align}
        \vec{CRB}_\mathrm{L}(\bm{\varphi}) = \mathbf{F}_\mathrm{u}(\bm{\varphi})^{-1} \preceq \vec{CRB}(\bm{\varphi}),
    \end{align}
    where 
    \begin{align}
        \mathbf{F}_\mathrm{u}(\bm{\varphi}) &= \frac{2}{\sigma^{2}} \Re\left\{\mathbf{U}_{0} \mathbf{U}_{0}^{H}\right\}+\frac{4}{\pi \sigma^{2}}\Re\left\{\mathbf{U}_{1} \mathbf{U}_{1}^{H}\right\} \nonumber \\
        & \triangleq \frac{2}{ \sigma^{2}}(\mathbf{U}_{\rm R}\bar{\mathbf {\Sigma}} \mathbf{U}_{\rm R}^T +\mathbf{U}_{\rm I}\bar{\mathbf {\Sigma}} \mathbf{U}_{\rm I}^T) ,
    \end{align}
    with 
    \begin{align}
        &\bar{\boldsymbol {\Sigma}} = \mathbf{I}_N\otimes \boldsymbol{\Sigma_0}, \nonumber \\
        &\boldsymbol{\Sigma}_0 = \left(1-\frac{2}{\pi}\right)\text{diag}(\boldsymbol{\delta} )+\frac{2}{\pi}\mathbf{I}_M.
    \end{align}

    The matrix $\bar{\boldsymbol {\Sigma}}$ is diagonal with its diagonal elements belonging to the binary set $\left\{\frac{2}{\pi}, 1\right\}$. Let $\bar{\mathbf{U}} = \mathbf{U}\bar{\boldsymbol{\Sigma}}^{1/2}$. We have 
    \begin{align}
        &\bar{\mathbf{U}}=[\bar{\boldsymbol{\Delta}}, \quad \bar{\mathbf{G}}, \quad \mathrm{i} \bar{\mathbf{G}}]^H,\\
        &\bar{\boldsymbol{\Delta}} = \bar{\boldsymbol {\Sigma}}^{1/2}\boldsymbol{\Delta} \label{eq:27}, 
        \\ &\bar{\mathbf{G}} =\mathbf{I}_N\otimes (\mathbf{\Sigma}_0^{1/2} \mathbf{A}). \label{eq:28}
    \end{align}
    With these notations, the upper bound of the FIM for the mixed-ADC data based model can be concisely written as
    \begin{align}
    \mathbf{F}_\mathrm{u}(\bm{\varphi})
    &=\frac{2}{\sigma^2} \Re\left\{\bar{\mathbf{U}} \bar{\mathbf{U}}^{H}\right\}.
    \end{align}

    To obtain the DOA-related block of the lower bound of the CRB matrix, we can compute it via block-wise matrix inversion \cite{stoica2005spectral}, i.e.,
    \begin{align} \label{eq:30}
        \vec{CRB}_\mathrm{L}(\bm{\omega})=\frac{\sigma^{2}}{2}\Re\left\{\bar{\boldsymbol\Delta}^{H} \boldsymbol\Pi_{\bar{\mathbf{G}}}^{\perp} \bar{\boldsymbol\Delta}\right\}^{-1},
    \end{align}
    where $\boldsymbol{\Pi}_{\bar{\mathbf{G}}}^{\perp}=\mathbf{I}-\bar{\mathbf{G}}\left(\bar{\mathbf{G}}^H \bar{\mathbf{G}}\right)^{-1} \bar{\mathbf{G}}^H$ is the orthogonal projector onto the null space of $\bar{\mathbf{G}}^H$ such that  $\bar{\mathbf{G}}^H\boldsymbol{\Pi}_{\bar{\mathbf{G}}}^{\perp} = \mathbf{0}$. Substituting (\ref{eq:27}) and (\ref{eq:28}) into (\ref{eq:30}), we obtain a clearer expression for $\vec{CRB}_\mathrm{L}(\bm{\omega})$:
    \begin{align}\label{eq:31}
    \vec{CRB}_\mathrm{L}(\bm{\omega})& =\frac{\sigma^{2}}{2}\Re\left\{\boldsymbol\Delta^{H}\bar{\boldsymbol {\Sigma}}^{\frac{1}{2}} \boldsymbol\Pi_{\bar{\mathbf{G}}}^{\perp} \bar{\boldsymbol {\Sigma}}^{\frac{1}{2}}\boldsymbol\Delta\right\}^{-1}
    \nonumber
    \\&\triangleq \frac{\sigma^2}{2N}\Re\{\left(\dot{\mathbf{A}}^H \boldsymbol{\Omega} \dot{\mathbf{A}}\right) \odot \hat{\mathbf{P}}^T\}^{-1},
    \end{align}
    where 
    \begin{align}
        \hat{\mathbf{P}}&=\frac{1}{N} \sum_{n=1}^N \mathbf{s}(n) \mathbf{s}^H(n) \label{eq_new:35},\\
        \boldsymbol{\Omega}&=\boldsymbol{\Sigma}_0-\boldsymbol{\Sigma}_0\mathbf{A}(\mathbf{A}^H\boldsymbol{\Sigma}_0\mathbf{A})^{-1}\mathbf{A}^H\boldsymbol{\Sigma}_0.
    \end{align}

    % The subsequent analysis aims to develop a more intuitive expression for the CRB of the DOA. 
    For the case of a single snapshot and a single target, the matrices $\mathbf{A}$ and $\dot{\vec{A}}$ are given by
    \begin{align}
        \mathbf{A}&=\begin{bmatrix} 1, e^{\mathrm{i}w},  \dots, e^{\mathrm{i}(M-1)w}\end{bmatrix}^T,
    \end{align}
    and 
    \begin{align}
        \dot{\vec{A}}&= \mathrm{i} \begin{bmatrix} 0, e^{\mathrm{i}w},  \dots, (M-1) e^{\mathrm{i} (M-1)w}\end{bmatrix}^T,
    \end{align}
    respectively. It follows that
    \small
    \begin{align}
        \mathbf{A}^H\bold\Sigma_0\mathbf{A} &= \sum_{i=1}^M\left[(1-\frac{2}{\pi})\delta_i+\frac{2}{\pi}\right] = M_0+\frac{2}{\pi}M_1, \\
        \dot{\mathbf{A}}^H\bold\Sigma_0\mathbf{A} &= -\mathrm{i}\left[\sum_{i=1}^M\delta_i (i-1)+\frac{2}{\pi}\sum_{i=1}^M\bar{\delta}_i (i-1)\right],\\
        \dot{\mathbf{A}}^H\boldsymbol{\Sigma}_0\dot{\mathbf{A}}& = \left[\sum_{i=1}^M\delta_i (i-1)^2+\frac{2}{\pi}\sum_{i=1}^M\bar{\delta}_i (i-1)^2\right].
    \end{align}
    \normalsize
    Inserting the above equations into the CRB expression in (\ref{eq:30}) , we obtain 
    \small
    \begin{align} \label{eq:39}
        \text{CRB}_\mathrm{L}(\omega)=\frac{\sigma^2(M_0+\frac{2}{\pi}M_1)}{2pS} = \frac{M_0+\frac{2}{\pi}M_1}{2S} \frac{1}{\text{SNR}},
    \end{align}
    \normalsize
    where $p$ is the signal power, $\text{SNR} =\frac{p}{\sigma^2}$ and $S$ is defined as
    \small
    \begin{align} \label{eq:40}
        S = &\left[\sum_{i=1}^M \delta_i (i-1)^2+\frac{2}{\pi} \sum_{i=1}^M \bar{\delta}_i (i-1)^2\right](M_0+\frac{2}{\pi}M_1) \nonumber\\
        &-\left[\sum_{i=1}^M \delta_i (i-1)+\frac{2}{\pi} \sum_{i=1}^M \bar{\delta}_i (i-1)\right]^2 \nonumber \\
         \triangleq & \sum_{i=1}^Mg_i (i-1)^2\sum_{i=1}^Mg_i-\left[\sum_{i=1}^Mg_i (i-1)\right]^2,
    \end{align}
    \normalsize
    where $g_i \in \left\{1, \frac{2}{\pi}\right\}$, $\sum_{i=1}^M g_i=M_0+\frac{2}{\pi}M_1$, \textcolor{black}{and $S$ is related to the placement of different types of ADC.} By using Lagrange’s identity, $S$ can be expressed as
    \begin{align}
        S = \sum_{i=1}\sum_{j>i}g_i g_j(j-i)^2.
    \end{align}

    % \subsection{Asymptotic CRB for DOA}
    For sufficiently large $N$,  $\hat{\mathbf{P}}$ in (\ref{eq_new:35}) is approximately equal to  \textcolor{black}{$\mathbf{P} = \lim_{N \to \infty}{\frac{1}{N}\sum_{n=1}^N\mathbf{s}(n)\mathbf{s}(n)^H} $.} Thus, the lower bound of CRB for mixed-ADC based DOA is given by
    \begin{align}
        \vec{CRB}_\mathrm{L}(\bm{\omega})  =\frac{\sigma^2}{2N}\Re\left\{\left(\dot{\mathbf{A}}^H \boldsymbol{\Omega} \dot{\mathbf{A}}\right) \odot \mathbf{P}^T\right\}^{-1}.
    \end{align}

    Consider the case that when $M$ increases, the number of high-precision ADCs $M_0$ does not increase proportionally with $M$. Motivated by (\ref{eq:39}), for computational simplicity, we can get the following asymptotic result (see Appendix \ref{Appendix:A} for the detailed derivations):
    \begin{align} \label{eq:43}
        \vec{CRB}_\mathrm{L}(\bm{\omega}) = \frac{M_0+\frac{2}{\pi}M_1}{2 NS}\left[\begin{array}{lll}
            \frac{1}{\text{SNR}_1} & & 0 \\
            & \ddots & \\
            0 & & \frac{1}{\text{SNR}_K}
            \end{array}\right],
    \end{align}
    where $\text{SNR}_i$ is the SNR of the $i$th signal. Thus, the asymptotic CRB for the $k$th target's DOA can be approximately expressed as:
    \begin{align} \label{eq:47}
        \text{CRB}_\mathrm{L}(\omega_k) =  \frac{M_0+\frac{2}{\pi}M_1}{2NS}\frac{1}{\text{SNR}_k}.
    \end{align}
    
    Consequently, given the power and angle of an incident signal, the asymptotic CRB for its  DOA depends solely on the number of high-precision ADCs, one-bit ADCs, and their placement in the array, \textcolor{black}{i.e., variable $S$.}
    For a given number of high-precision and one-bit ADCs, i.e., $M_0$ and $M_1$, respectively, our interest is to minimize the  asymptotic CRB (i.e., maximizing $S$). Then the optimization problem can be formulated as
\small
\begin{align}
    &\max_{\{g_i\}_{i=1,2\cdots,M}} \quad S=\sum_{i=1}^M\sum_{j>i}g_i g_j(j-i)^2 \nonumber \\ 
    & \ \ {\rm s.t.} \quad  g_i \in\left\{1, \frac{2}{\pi}\right\}, \quad i=1,2,\dots,M, \nonumber\\
    &  \qquad \quad \  \sum_{i=1}^M g_i=M_0+\frac{2}{\pi}M_1,  \label{eq:45} 
\end{align}
\normalsize
which is referred to as the ADC placement optimization for the mixed-ADC based architecture.

\subsubsection{More high-precision ADCs yielding better asymptotic CRB} \label{sec:5.1}
    Suppose that the total number of ADC pairs is fixed as $M$. Let $\{\tilde{g}_i\}_{i=1}^M$ denote the replacement of the $k$th  ADC pair from one-bit to high-precision by altering $\{g_i\}_{i=1}^M$, which originally has $M_0-1$ pairs of high-precision ADCs.  We define $F$ and $\tilde{F}$ as:
    \begin{align}
        F &= \frac{\sum_{i=1}^M\sum_{j>i}g_i g_j(j-i)^2}{M_0-1+\frac{2}{\pi}(M_1+1)},
    \end{align}
    and 
    \begin{align}
        \tilde{F} &= \frac{\sum_{i=1}^M\sum_{j>i}\tilde{g}_i \tilde{g}_j(j-i)^2}{M_0+\frac{2}{\pi}M_1},
    \end{align}
    respectively.
    Note that 
    \begin{align}
        \sum_{i=1}\sum_{j>i}g_i g_j(j-i)^2 = g_k\sum_{j=1, j\neq k}^Mg_j(j-k)^2+C,
    \end{align}
    where $C$ is a constant independent of $g_k$. Then we have
    \small
    \begin{align}\label{eq:49}
        \tilde{F} - F = \frac{(1-\frac{2}{\pi})\{(M_0+\frac{2}{\pi}M_1-1)\sum_{j=1,j\neq k}^Mg_j(j-k)^2 - C\}}{(M_0+\frac{2}{\pi}M_1)(M_0-1+\frac{2}{\pi}(M_1+1))}.
    \end{align}
    \normalsize
    In Equation (\ref{eq:49}), we use $\{\tilde{g}_i\}_{i=1}^M$ in lieu of $\{g_i\}_{i=1}^M$ since $g_k$ has no influence on the overall result. Then we have:
    \small
    \begin{align} \label{eq:50}
        &(M_0+\frac{2}{\pi}M_1-1)\sum_{j=1,j\neq k}^Mg_j(j-k)^2 - C
        \nonumber\\ 
        &= \sum_{j=1}^M\tilde{g}_j\sum_{j=1}^M\tilde{g}_j(k-j)^2-\sum_{i=1}^M\sum_{j>i}\tilde{g}_i\tilde{g}_j(j-i)^2 \nonumber
        \\&\geq -\left(\sum_{j=1}^Mj\tilde{g}_j\right)^2+\left(\sum_{j=1}^M\tilde{g}_jj^2\right)\left(\sum_{j=1}^M\tilde{g}_j\right)-\sum_{i}\sum_{j>i}\tilde{g}_i\tilde{g}_j(j-i)^2 \nonumber
        \\& =\sum_i\sum_j \tilde{g}_i\tilde{g}_j j^2 - \frac{1}{2}\sum_i\sum_j\tilde{g}_i\tilde{g}_j(i^2+j^2) =0,
    \end{align}
    \normalsize
    where the inequality in (\ref{eq:50}) follows from the property of the quadratic function, and the equality holds when $k = \frac{\sum_{j=1}^M\tilde{g}_jj}{\sum_{j=1}^M\tilde{g}_j}$. 
    
    Thus, we have  $\tilde{F} \geq F$, indicating that for the fixed $M$, the asymptotic CRB decreases monotonically as the number of high-precision ADCs $M_0$ increases.

\subsubsection{Upper and lower bounds of the asymptotic CRB of the mixed-ADC architecture}
When the ULA is equipped with only high-precision ADCs, the asymptotic CRB of the $k$th DOA in (\ref{eq:47}) becomes 
\begin{align}
    \text{CRB}_{\mathrm{L},0}(\omega_k) = \frac{6}{NM(M^2-1)}\frac{1}{\text{SNR}_k},
\end{align}
which is consistent with the result provided in \cite{stoica1989music}. On the other hand, when the ULA is equipped with only one-bit ADCs, the upper bound of the asymptotic CRB of the $k$th DOA is given by:
\begin{align}
    \text{CRB}_{\mathrm{L},1}(\omega_k) = \frac{\pi}{2}\frac{6}{NM(M^2-1)}\frac{1}{\text{SNR}_k}.
\end{align}
Thus, the asymptotic CRB for the $k$th DOA under the mixed-ADC architecture is bounded as follows:
\small
\begin{align}
    \frac{6}{NM(M^2-1)}\frac{1}{\text{SNR}_k} \leq \text{CRB}_{\mathrm{L},\mathrm{m}}(\omega_k) \leq \frac{3\pi}{NM(M^2-1)}\frac{1}{\text{SNR}_k}.
\end{align}
\normalsize
Compared to the purely high-precision ADC system, the purely one-bit ADC system has a $1.96$ dB  loss on the asymptotic CRB of each DOA.

\subsection{Placement of High-precision ADCs}
To solve the problem in (\ref{eq:45}), we propose a strategy of swapping the positions of the high-precision and one-bit ADCs to obtain the optimal placement of high-precision ADCs for a fixed $M_0$ and $M_1$. Specifically, let $\tilde{S}(\tilde{g}_m = 1, \tilde{g}_n = \frac{2}{\pi},\dots)$ denote the placement obtained by swapping the $m$th and $n$th ADCs in the original placement $S(g_m = \frac{2}{\pi}, g_n = 1,\dots)$. We consider following two specific situations to swap the positions of the $m$th and $n$th ADC pairs:
\begin{enumerate}
    \item $m \leq M_h$, where $M_\mathrm{h} = \lfloor \frac{M_0+1}{2} \rfloor$. In this case, $m$ corresponds to the index of the smallest number among the one-bit ADC pairs. Additionally, we choose $n$ corresponding to the index of the smallest number while larger than $m$ among the high-precision ADC pairs.
    \item $m \geq M - M_\mathrm{h} + 1$, where $m$ corresponds to the index of the largest number among the one-bit ADC pairs. We choose $n$ corresponding to the index of the largest number while smaller than $m$ among the high-precision ADC pairs.
\end{enumerate}

Note that $S$  can be expressed as
\small
\begin{align}
    &S = g_m\sum_{j=1, j\neq n}^Mg_j(j-m)^2 + g_n\sum_{j=1, j\neq m}^Mg_j(j-n)^2 \nonumber\\
    &+ g_mg_n(m-n)^2+C ,
    % &\tilde{S} = \tilde{g}_m\sum_{j=1, j\neq m}^M\tilde{g}_j(j-m)^2 + \tilde{g}_n\sum_{j=1, j\neq n}^M\tilde{g}_j(j-n)^2 \nonumber \\
    % &+ \tilde{g}_m\tilde{g}_n(m-n)^2+C,
\end{align}
\normalsize
and $\tilde{S}$  is similar, where the constant $C$ is not related to $m$ or $n$. Then we have
\small
\begin{align}\label{eq:56}
    \tilde{S} - S = \left(1-\frac{2}{\pi}\right)(n-m)\sum_{j=1, j\neq m,n}^Mg_j(2j-m-n).
\end{align}
\normalsize

Since  $\tilde{S}-S >0$ in both cases (see Appendix \ref{Appendix:B} for details), we have the flexibility to iteratively adjust the positions of the high-precision and one-bit ADC pairs  in order to incrementally increase $S$  until none of the aforementioned situations remains. As a result, the high-precision ADCs are evenly distributed around the edges of the array, as shown in Fig. 1(c), which represents the optimal ADC placement of a mixed-ADC base architecture.

\remark The above conclusion can be extended to a more general case where the ULA is equipped with ADCs of any two different quantization precisions. In order to achieve best performance, the higher precision ADCs should be evenly distributed around the edges of the ULA.

It is demonstrated in \cite{9664619} that the CRB decreases as the quantization becomes finer, which indicates that the coefficients (i.e., $\rho_1=\frac{2}{\pi}$ in the case of one-bit quantization) associated with higher quantization bits tend to be larger. Let $\Gamma$ denote the set of coefficients as 
\begin{align}
    \Gamma \triangleq \left\{\frac{2}{\pi} = \rho_1 < \rho_2 < \dots <\rho_l< \rho_0=1\right\},
\end{align}
where $\rho_i$ denotes the coefficient of $i$-bit quantization and $\rho_0$ denotes the coefficient of high-precision. For a system with ADCs of two different quantization precisions, characterized by $p$-bit and $q$-bit ($p<q$), the corresponding coefficients and numbers are $\rho_p$, $\rho_q$, $M_p$ and $M_q$. Then the  optimization problem can be reformulated as
\small
\begin{align}
    &\max_{\{g_i\}_{i=1,2\cdots,M}} \quad S=\sum_{i=1}^M\sum_{j>i}g_i g_j(j-i)^2 \nonumber \\ 
    & \ \ {\rm s.t.} \quad  g_i \in \left\{\rho_p, \rho_q\right\}, \quad i=1,2,\dots,M \nonumber \\
     &  \qquad \quad \  \sum_{i=1}^M g_i=M_p\rho_p+M_q\rho_q . 
\end{align}
\normalsize
Based on the symmetry of the ULA and the analysis in Appendix \ref{Appendix:B}, we consider the  case that $m \leq \lfloor \frac{M_q+1}{2} \rfloor$, where $m$ corresponds to the index of the smallest number among the ADC pairs with a lower quantization precision, and $n$ corresponds to the index of the largest number among the ADC pairs with a higher quantization precision.

Let $\tilde{S}(\tilde{g}_m = \rho_p, \tilde{g}_n = \rho_q,\dots)$ denote the placement obtained by swapping the positions of the $m$th and $n$th ADC pairs from original placement $S(g_m = \rho_q, g_n = \rho_p,\dots)$. Using a similar method in (\ref{eq:56}), we have
\begin{align}
    \tilde{S} - S  &= (\rho_p - \rho_q)(n-m)\sum_{j=1, j\neq m,n}^Mg_j(2j-m-n) \nonumber \\
    & = (\rho_p - \rho_q)(n-m)H(m,n) >0
\end{align}

Thus, the ADC pairs with a higher precision should be evenly distributed around the edges of the ULA for the best performance under mixed-ADC architecture.

 \textcolor{black}{\remark Different mixed-ADC placements can also significantly impact the performance of non-ULA geometries, such as sparse linear arrays (SLA), uniform rectangular arrays, uniform circular arrays, L-shape arrays, etc. Due to their unique geometries, the optimal placements of mixed-ADCs in these scenarios differ from that for ULAs. They can be subjects of further studies following our framework. For example, for the case of SLAs, this problem can be transformed into a 0–1 integer quadratic programming problem, as analyzed in \cite{zhang2023sla}. Also, for arrays with symmetrical properties, the swapping-based method remains an effective tool for achieving globally optimal solutions.}.

\subsection{Performance efficiency}

We propose a method to analyze the relationship between the proportion of high-precision ADCs and system performance under mixed-ADC architecture with the optimal mixed-ADC placement. This is accomplished by introducing the performance efficiency factor $\eta_{\text{PF}}$ for a mixed-ADC architecture, which is given by 
\begin{align}
    \eta_{\text{PF}} = \frac{\text{CRB}_{\rm{HP}}}{\text{CRB}_{\kappa}}.
\end{align}
Here $\text{CRB}_{\rm{HP}}$ and $\text{CRB}_{\kappa}$ represent the CRB of high-precision system and the mixed-ADC architecture with a $\kappa$  proportion of high-precision ADCs, respectively.

Notably, a higher value of $\eta_{\text{PF}}$ indicates a superior estimation performance achieved by the mixed-ADC architecture. When the proportion $\kappa = 1$, representing a high-precision system, the performance efficiency is equal to 1.

\section{Parameter Estimation for Mixed-ADC System}\label{sec:6}

Due to the asymptotic efficiency of the ML estimators, we consider ML parameter estimation for mixed-ADC systems. The negative log-likelihood function for the data model in (\ref{eq:5}) is given as
\small
\begin{align} 
    &-\ln{L}(\boldsymbol{\chi})=- \ln{L}_1(\boldsymbol{\chi})-\ln{L}_0(\boldsymbol{\chi}) \nonumber \\
    =&-\sum_{n=1}^N \sum_{m=1}^{M_1} \ln \left(\Phi \left(Y_{1,\rm R}(m,n) \frac{\Re\{\mathbf{a}_{1,m}^T\mathbf{s}(n)\}-H_{1,\rm R}(m,n)}{\sigma/\sqrt{2}}\right)\right) \nonumber \\
    &-\sum_{n=1}^N \sum_{m=1}^{M_1} \ln \left(\Phi \left(Y_{1,\rm I}(m,n) \frac{\Im\{\mathbf{a}_{1,m}^T\mathbf{s}(n)\}-H_{1,\rm I}(m,n)}{\sigma/\sqrt{2}}\right)\right) \nonumber \\
    & + \frac{1}{\sigma^2}||\mathbf{Y_0}-\mathbf{A}_0\mathbf{S}||_{\rm F}^2+M_0N\ln{\sigma^2}+M_0N\ln{\pi},
\end{align}
\normalsize
where $\mathbf{a}_{1,m}$ denotes the $m$th column of $\mathbf{A}_1^T$. For notational simplicity, let $\zeta = \sqrt{2}/\sigma$, $\mathbf{b}(n) = \zeta\mathbf{s}(n)$ and $\mathbf{B} = [\mathbf{b}(1), \dots, \mathbf{b}(N)]$. Then the negative log-likelihood function can be simplified as
\begin{align} 
    &-\ln{L}(\tilde{\bm{\chi}})  \nonumber \\
    =&-\sum_{n=1}^N \sum_{m=1}^{M_1} \ln \left(\Phi \left(Y_{1,\rm R}(m,n) (\Re\{\mathbf{a}_{1,m}^T\mathbf{b}(n)\}-\zeta H_{1,\rm R}(m,n)\right)\right) \nonumber \\
    &-\sum_{n=1}^N \sum_{m=1}^{M_1} \ln \left(\Phi \left(Y_{1,\rm I}(m,n) (\Im\{\mathbf{a}_{1,m}^T\mathbf{b}(n)\}-\zeta H_{1,\rm I}(m,n)\right)\right) \nonumber \\
    & + \frac{1}{2}||\zeta \mathbf{Y_0}-\mathbf{A}_0\mathbf{B}||_{\rm F}^2-M_0N\ln{\zeta^2}+M_0N\ln{2\pi},
\end{align}
where $\tilde{\bm{\chi}} = [\bm{\omega}^T, \mathbf{b}^T_{\rm R}, \mathbf{b}^T_{\rm I}, \zeta]^T$.

It should be noted that the negative log-likelihood function above is a  highly nonlinear and non-convex function of $\bm{\omega}$. Consequently, global optimization of this function is rather difficult and challenging. However, it is worth mentioning that the optimization problem becomes convex when considering the variable vector $\{\mathbf{b}^T_{\rm R}, \mathbf{b}^T_{\rm I}, \zeta\}^T$, thereby enabling efficient solutions using a convex optimization method, e.g., the Newton's method. Conventionally, the ML estimator is implemented as follows:
\begin{enumerate}
    \item Conduct a $K$-dimensional exhaustive coarse search on the angular spaces to find coarse estimates of $\bm{\omega}$.\label{step1}
    \item Determine the optimal estimates of $\mathbf{B}$ and $\zeta$ given $\bm{\omega}$.
\end{enumerate}

When considering a grid based method, the angle space is partitioned into $K_{\omega}$ points, resulting in a computational complexity of $\mathcal{O}(K_{\omega}^K)$. As the number of targets increases, the complexity of the ML estimator above becomes computationally prohibitive and time-consuming. 

To overcome this issue, we propose an efficient hybrid algorithm to realize the ML estimation. Initially, we partition the angular space into $K_{\omega}$ points, and modify the SLIM algorithm \cite{tan2010sparse} for the mixed-ADC architecture. Subsequently, we use RELAX \cite{li1996efficient} to cyclically refine the parameters of each target initialized by the SLIM estimates to obtain the ML estimates.

\subsection{SLIM for Initialized Parameter Estimation}

We divide the continuous angular space uniformly into $K_{\omega}$ grid points. For a sufficiently fine grid, we consider the following Bayesian model \cite{tan2010sparse}:
\small
\begin{align}
    p(\mathbf{Y}|\mathbf{B},\zeta) = &\prod_{n=1}^N\prod_{m=1}^{M_1}\Phi \left(Y_{1,\rm R}(m,n) (\Re\{\mathbf{a}_{1,m}^T\mathbf{b}(n)\}-\zeta H_{1,\rm R}(m,n)\right) \nonumber\\
    &\Phi \left(Y_{1,\rm I}(m,n) (\Im\{\mathbf{a}_{1,m}^T\mathbf{b}(n)\}-\zeta H_{1,\rm I}(m,n)\right) \nonumber\\ &\times \left(\frac{\zeta^2}{4\pi}\right)^{M_0N}e^{-\frac{1}{2}\|\zeta \mathbf{Y_0}-\mathbf{A}_0\mathbf{B}\|_{\rm F}^2}, \nonumber \\
    p(\mathbf{B})  \propto&  \prod_{r=1}^{K_{\omega}}e^{-\frac{2}{q}(\|\mathbf{b}_r\|_2^q-1)}, \quad p(\zeta)  \propto 1,
\end{align}
\normalsize
where $\mathbf{b}_r$ denotes the $r$th column of $\mathbf{B}^T$,  $p(\mathbf{B})  \propto \prod_{r=1}^{K_{\omega}}e^{-\frac{2}{q}(\|\mathbf{b}_r\|_2^q-1)}$ is a sparsity promoting prior for $0 < q \leq 1$ performed on each row of $\mathbf{B}$ to exploit the joint sparsity, and $p(\zeta)$ is a noninformative prior.
When $q \to 0$, the sparsity promoting term becomes $p(\mathbf{B}) \propto \prod_{r=1}^{K_\theta}\frac{1}{\|\mathbf{b}_r\|_2^2}$.
To estimate $\mathbf{B}$ and $\zeta$, we adopt the \textcolor{black}{maximum a posteriori} approach:
\begin{align}
    (\hat{\mathbf{B}}, \hat{\zeta}) =  \arg \max _{\mathbf{B}, \zeta} p(\mathbf{Y} | \mathbf{B}, \zeta) p(\mathbf{B}) p(\zeta).
\end{align}

By taking the negative logarithm and ignoring the constant terms, the above problem can be written as
\small
\begin{align}\label{eq:61}
    \min _{\mathbf{B}, \zeta} g(\mathbf{B}, \zeta) \triangleq l_1(\mathbf{B}, \zeta) + l_0(\mathbf{B}, \zeta) + \sum_{r=1}^{K_{\omega}}\frac{2}{q}(\|\mathbf{b}_r\|_2^q-1),  
\end{align}
\normalsize
where $l_1(\mathbf{B}, \zeta) = -\ln{L}_1(\mathbf{B}, \zeta)$ and $l_0(\mathbf{B}, \zeta) = \frac{1}{2}\|\zeta \mathbf{Y_0}-\mathbf{A}_0\mathbf{B}\|_{\rm F}^2-M_0N\ln{\zeta^2}$ can be seen as data fitting terms for one-bit and high-precision data, respectively.

To efficiently minimize the complicated and non-convex objective function in (\ref{eq:61}), we employ the majorization-minimization (MM) technique to attain enhanced computational efficiency. \textcolor{black}{Specifically, given the estimates of $\mathbf{B}$ and $\zeta$ at the $i$th MM iteration, the majorizing function for $g(\mathbf{B},\zeta)$, denoted as $G(\mathbf{B},\zeta|\hat{\mathbf{B}}^i,\hat{\zeta}^i)$, can be constructed as follows:
\small
\begin{align}\label{eq:68}
    &G(\mathbf{B},\zeta|\hat{\mathbf{B}}^i,\hat{\zeta}^i) =  \frac{1}{2}\|\mathbf{A}_1\mathbf{B} - (\zeta \mathbf{H}+\hat{\mathbf{D}}^i)\|_{\rm F}^2 \nonumber \\&+ \frac{1}{2}\|\zeta\mathbf{Y}_0-\mathbf{A}_0\mathbf{B}\|_{\rm F}^2+\frac{1}{N}\Tr(\mathbf{B}^H(\hat{\mathbf{P}}^i)^{-1}\mathbf{B})-2M_0N\ln{\zeta},
\end{align}
\normalsize
where $\hat{\mathbf{D}}^i$ and $\hat{\mathbf{P}}^i$ are defined using the estimates obtained from the $i$th MM iteration (see Appendix \ref{Appendix:C} for details).}

The minimization of the function (\ref{eq:68}) can be achieved by using a cyclic algorithm \cite{stoica2004cyclic}, which alternates between the minimization of $G(\mathbf{B},\zeta|\hat{\mathbf{B}}^i,\hat{\zeta}^i)$ with respect to $\zeta$ for fixed $\mathbf{B}$ and with respect to $\mathbf{B}$ for given $\zeta$. Note that $G(\mathbf{B},\zeta|\hat{\mathbf{B}}^i,\hat{\zeta}^i)$ with respect to $\mathbf{B}$ and $\zeta$ is an unconstrained quadratic optimization problem, it can be minimized by  setting the complex derivatives $(d/d\mathbf{B}^H)G(\mathbf{B},\zeta|\hat{\mathbf{B}}^i,\hat{\zeta}^i)$ and $(d/d\zeta)G(\mathbf{B},\zeta|\hat{\mathbf{B}}^i,\hat{\zeta}^i)$ to zero cyclicly to obtain $\hat{\mathbf{B}}^{i+1}$ and $\hat{\zeta}^{i+1}$. 

To update $\mathbf{B}$ based on the estimates obtained during the $t$th inner cyclic iteration, we introduce a new variable $\tilde{\mathbf{Y}}$ with $\tilde{\mathbf{Y}}(\bm{\delta},:) = \zeta\mathbf{Y}_0$ and $\tilde{\mathbf{Y}}(\bm{\bar{\delta}},:) = \zeta \mathbf{H}+\hat{\mathbf{D}}^{i+1}_t$. By ignoring the irrelevant variable, the optimization objective function can be written as 
\small
\begin{align}\label{eq:73}
    G(\mathbf{B},\zeta|\hat{\mathbf{B}}^{i+1}_t,\hat{\zeta}^{i+1}_t) = \frac{1}{2}\|\tilde{\mathbf{Y}}-\mathbf{A}\mathbf{B}\|_{\rm F}^2+\frac{1}{N}\Tr(\mathbf{B}^H(\hat{\mathbf{P}}^{i+1}_t)^{-1}\mathbf{B}).
\end{align}
\normalsize
Setting the derivative of (\ref{eq:73}) with respect to $\mathbf{B}$ to zero leads to
\begin{align}
     (\mathbf{A}^H\mathbf{A}+\frac{2}{N}(\mathbf{P}^{i+1}_t)^{-1})\mathbf{B} = \mathbf{A}\tilde{\mathbf{Y}}.
\end{align}
Since  $\mathbf{A}^H\mathbf{A}+\frac{2}{N}(\mathbf{P}^{i+1}_t)^{-1}$ is a positive definite matrix, we can obtain $\mathbf{B}$ as
\begin{align} \label{eq:78}
    \mathbf{B} = \mathbf{P}_{t}^{i+1}\mathbf{A}^H( \mathbf{R}_{t}^{i+1})^{-1}\tilde{\mathbf{Y}},
\end{align}
where 
\begin{align} \label{eq:79}
    \hat{\mathbf{R}}_t^{i+1} = \mathbf{A}\mathbf{P}^{i+1}_t\mathbf{A}^H+\frac{2}{N}\mathbf{I}_M.
\end{align}

For $\hat{\zeta}_{t+1}^{i+1}$, let $\mathbf{T}$ and $\mathbf{Q}$ be the new variables with $\mathbf{T}[\bm{\delta},:] = \mathbf{Y}_0$, $\mathbf{T}[\bm{\bar{\delta}},:] = \mathbf{H}$, $\mathbf{Q}[\bm{\delta},:] = \bm{0}$ and $\mathbf{Q}[\bm{\bar{\delta}},:] = \hat{\mathbf{D}}_t^{i+1}$. Setting the derivative of (\ref{eq:68}) with respect to $\zeta$ to zero yields:
\begin{align}\label{eq:80}
    \Tr(\mathbf{T}^H\mathbf{T})\zeta - \Re\{\Tr(\mathbf{T}^H(\mathbf{A}\mathbf{B} - \mathbf{Q}))\}-\frac{2M_0N}{\zeta} = 0.
\end{align}
Substituting $\mathbf{B}$ from (\ref{eq:78}) into (\ref{eq:80}), we obtain the following quadratic function:
% \small
\begin{align}\label{eq:81}
    % \Tr(\mathbf{T}^H(\hat{\mathbf{R}}^{i+1}_t)^{-1}\mathbf{T})\zeta^2+\Re\{\Tr(\mathbf{T}^H(\hat{\mathbf{R}}^{i+1}_t)^{-1}\mathbf{Q})\}\zeta-2M_0N = 0.
        v\zeta^2+u\zeta-2M_0N = 0.
\end{align}
% \normalsize
where $u = \Re\{\Tr(\mathbf{T}^H(\hat{\mathbf{R}}^{i+1}_t)^{-1}\mathbf{Q})\}$ and $v =\Tr(\mathbf{T}^H(\hat{\mathbf{R}}^{i+1}_t)^{-1}\mathbf{T})$. Considering the fact that $\zeta \geq 0$, the solution can be obtained by the larger root of (\ref{eq:81}):
\begin{align}\label{eq:82}
    \hat{\zeta}_{t+1}^{i+1} = \frac{-u+\sqrt{u^2+8vM_0N}}{2v}.
\end{align}
Finally, substituting $\hat{\zeta}_{t+1}^{i+1}$ into (\ref{eq:78}) with $\tilde{\mathbf{Y}}$, we get:
\begin{align}\label{eq:83}
    \mathbf{B}_{t+1}^{i+1} = \mathbf{P}_{t}^{i+1}\mathbf{A}^H( \mathbf{R}_{t}^{i+1})^{-1}\tilde{\mathbf{Y}}_{t+1}^{i+1}.
\end{align}

In the inner cyclic iteration, we update $\mathbf{B}$ and $\zeta$ alternatingly until convergence. The majorizing function $G(\mathbf{B}, \zeta)$ is guaranteed to decrease monotonically since the following property of cyclic algorithm holds:
\begin{align}
    G(\hat{\mathbf{B}}_t^{i+1},\hat{\zeta}_t^{i+1}|\hat{\mathbf{B}}^i,\hat{\zeta}^i) &\geq G(\hat{\mathbf{B}}_t^{i+1},\hat{\zeta}_{t+1}^{i+1}|\hat{\mathbf{B}}^i,\hat{\zeta}^i) \nonumber \\
    &\geq G(\hat{\mathbf{B}}_{t+1}^{i+1},\hat{\zeta}_{t+1}^{i+1}|\hat{\mathbf{B}}^i,\hat{\zeta}^i).
\end{align}

By making use of the MM technique, the so-derived majorizing function is much easier to minimize than the original objective function. 
\textcolor{black}{The objective function presented in (\ref{eq:61}) is reduced by the SLIM algorithm, converging to a local minimum. The monotonicity of the SLIM algorithm is guaranteed, as shown by the following inequalities:
\begin{align}
&g(\mathbf{B}^{i+1}, \zeta^{i+1}) \leq G(\mathbf{B}^{i+1}, \zeta^{i+1}|\mathbf{B}^{i}, \zeta^{i}) \nonumber\\ &\leq G(\mathbf{B}^{i}, \zeta^{i}|\mathbf{B}^{i}, \zeta^{i}) = g(\mathbf{B}^{i}, \zeta^{i}),
\end{align}
where the first inequality and the final equality are the consequences of the majorizing functions' properties. The second inequality is a result of the cyclic minimization.}

\textcolor{black}{Finally, since the continuous angular space is uniformly divided into $K_{\omega}$ grid points, the DOA power spectrum is calculated by 
\begin{align}
    p_r=\frac{1}{N}\|\hat{\mathbf{s}}_r\|_2^{2-q} \quad r= 1,2,...,K_{\omega}.
\end{align}
where $\hat{\mathbf{s}}_r$ denotes the $r$th column of $\hat{\mathbf{S}}$ with $\hat{\mathbf{S}} = \hat{\mathbf{B}}/\hat{\zeta}$. We summarize the steps of the extended SLIM Algorithm for the mixed-ADC architecture in Algorithm \ref{alg:1}.}
\begin{algorithm}
	%\textsl{}\setstretch{1.8}
	\renewcommand{\algorithmicrequire}{\textbf{Input:}}
	\renewcommand{\algorithmicensure}{\textbf{Output:}}
	\caption{The extended SLIM Algorithm}
	\label{alg:1}
	\begin{algorithmic}[1]
	    \REQUIRE \textcolor{black}{The mixed-ADC output: $\mathbf{Y}$, including high-precision output:$\mathbf{Y}_0$ and one-bit output: $\mathbf{Y}_1$}, one-bit threshold: $\mathbf{H}$, the dictionary matrix: $\mathbf{A}$, high-precision ADC index: $\bm{\delta}$ and one-bit ADC index: $\bm{\bar{\delta}}$, outer threshold: $\varepsilon_1$, inner threshold: $\varepsilon_2$
		\STATE Initialize  $\hat{\mathbf{B}}^0$ and $\hat{\zeta}^0 > 0$, $i = 0$
		\STATE Construct $\mathbf{T}$ with $\mathbf{T}(\bm{\bar{\delta}},:)=\mathbf{H}$ and $\mathbf{T}(\bm{\delta},:)=\mathbf{Y}_0$
		\REPEAT
		\STATE Construct $\hat{\mathbf{D}}^i$ and $\xi^i$ using (\ref{eq:72})
		\STATE \textcolor{black}{$\hat{\mathbf{B}}_{0}^{i+1} = \hat{\mathbf{B}}^i, \hat{\zeta}^{i+1}_{0} = \hat{\zeta}^i, t= 0$}
		\REPEAT
		\STATE Construct $\hat{\mathbf{Q}}^{i+1}_{t+1}$ with $\hat{\mathbf{Q}}^{i+1}_{t+1}(\bm{\bar{\delta}},:)=\hat{\mathbf{D}}^{i+1}_{t+1}$ and $\hat{\mathbf{Q}}^{i+1}_{t+1}(\bm{\delta},:)=\bm{0}$
% 		\STATE Construct $\tilde{\mathbf{Y}}$ with $\hat{\mathbf{B}}_{t+1}^i$ and $\hat{\zeta}_{t+1}^i$ using (\ref{eq:71})
        \STATE Construct $\hat{\mathbf{P}}_{t+1}^{i+1}$ with $\hat{\mathbf{B}}_{t}^{i+1}$ using (\ref{eq:74})
		\STATE Construct $\hat{\mathbf{R}}_{t+1}^{i+1}$ using (\ref{eq:79})
		\STATE Update $\hat{\zeta}_{t+1}^{i+1}$ using (\ref{eq:82})
		\STATE Update $\hat{\mathbf{B}}_{t+1}^{i+1}$ using (\ref{eq:83})
            \STATE \textcolor{black}{Update $\hat{\mathbf{D}}_{t+1}^{i+1}$ using (\ref{eq:72})}
		\STATE $t = t + 1$
		\UNTIL practical convergence $\left(\|\hat{\mathbf{P}}_{t+1}^{i+1}-\hat{\mathbf{P}}_{t}^{i+1}\|_{\text{F}} / \|\hat{\mathbf{P}}_{t}^{i+1}\|_{\text{F}} < \varepsilon_2\right)$
		\STATE $i = i + 1$
		\UNTIL practical convergence $\left(\|\hat{\mathbf{P}}^{i+1}-\hat{\mathbf{P}}^{i}\|_{\text{F}} / \|\hat{\mathbf{P}}^{i}\|_{\text{F}} < \varepsilon_1\right)$
		\STATE Correct data $\hat{\mathbf{S}} = \hat{\mathbf{B}}/\hat{\zeta}$ and $\hat{\sigma} = \sqrt{2}/\hat{\zeta}$
                \STATE \textcolor{black}{Calculate DOA power spectrum $p_r=\frac{1}{N}\|\hat{\mathbf{s}}_r\|_2^{2-q} \quad r= 1,2,...,K_{\omega}$}
		\ENSURE  Sparse signal matrix $\hat{\mathbf{S}}$, noise variance $\hat{\sigma}$ \textcolor{black}{and DOA power spectrum $\{p_1,...,p_{K_{\omega}}\}$}
	\end{algorithmic}  
\end{algorithm}}

\begin{algorithm}
	%\textsl{}\setstretch{1.8}
	\renewcommand{\algorithmicrequire}{\textbf{Input:}}
	\renewcommand{\algorithmicensure}{\textbf{Output:}}
	\caption{SLIM-RELAX with mBIC}
	\label{alg:2}
	\begin{algorithmic}[1]
	    \REQUIRE Maximum model order: $K_{\text{max}}$, the coarse estimations of $K_{\text{max}}$ signal from SLIM 
		\FOR{$K = 1$ to $K_{\text{max}}$}
		\STATE Initialize $K$ signal from top $K$ estimates of SLIM.
		\REPEAT
		\STATE \{$\hat{\bm{\psi}}_1, \hat{\sigma}\} = \arg \min _{\bm{\psi}_1, \sigma}-\ln{L}(\bm{\psi}_1, \sigma| \{\bm{\psi}_j\}_{j=2}^{K})$ 
		\FOR{$k=2$ to $K$}
		\STATE $\hat{\bm{\psi}}_k = \arg \min _{\bm{\psi}_k}-\ln{L}(\bm{\psi}_k| \{\bm{\psi}_j\}_{j=1, j\neq k}^{K})$ 
		\ENDFOR
		\UNTIL practical convergence
		\STATE $\text{mBIC}(K)= -2 \ln{L}(\bm{\chi}) +(2 K N+3K) \ln{MN}$
		\ENDFOR
		
		\STATE $\hat{K} = \arg \min _K \text{mBIC}(K)$
		\ENSURE Estimated signal number $\hat{K}$ and corresponding parameter $\bm{\psi}$
	\end{algorithmic}  
\end{algorithm}

\subsection{SLIM-RELAX}
% \subsection{Refine the parameters with RELAX}

% When the number of targets, i.e., the true model order $K$, is known, we can use  RELAX algorithm to refine the coarse estimation of angle and signal waveform, which uses the idea of cyclic optimization to estimate the parameters of each signal, and only estimates the parameters of the current signal in each cyclic optimization step. Specifically, we denote $\mathcal{P} = \{\bm{\psi}_1, \bm{\psi}_2,...,\bm{\psi}_K\}$ as the parameter of the $K$ signal, where $\bm{\psi}_k = \{\theta_k, \mathbf{s}_k(1),\dots,\mathbf{s}_k(N)\}$ is the parameter set of the $k$th target. We select the top $K$ maximum power peak and corresponding parameter from sparse estimate of MMSLIM as the coarse estimates of the $K$ source signal.

When the number of targets, i.e., the true model order $K$, is known, we can use the RELAX algorithm \cite{li1996efficient} to estimate the parameters of each signal cyclically to refine the coarse parameter estimates. 

Specifically, we denote by $\bm{\psi}_k = \{\theta_k, \mathbf{s}^T_{k, \rm R}, \mathbf{s}^T_{k, \rm I}\}$ as the parameter vector corresponding to the $k$th target, where $\mathbf{s}_k$ is the $k$th column of $\mathbf{S}^T$. Let   $\mathcal{P} = \{\hat{\bm{\psi}}_1, \hat{\bm{\psi}}_2,...,\hat{\bm{\psi}}_K\}$, where $\hat{\bm{\psi}}_k$ is the coarsely estimated parameter vector corresponding to the $k$th largest signal from the SLIM.    

In the cyclic optimization step, we refine the parameter vector of the $k$th signal  while fixing the parameters of the other signals, i.e., $\{\bm{\psi}_j\}_{j=1, j\neq k}^{K}$. For noise power $\sigma^2$, we update it along with the first target parameter set $\bm{\psi}_1$. \textcolor{black}{Given that SLIM is a grid-based algorithm, the error in the initial estimates us constrained within $\frac{2\pi}{K_{\omega}}$. This limitation results in a multi-variable non-convex problem with bounded variables.}. Thus, the refinement can be performed by using the interior-point based bounded optimization method (e.g., the ``fmincon'' of MATLAB) over the angular interval $\left[\hat{\omega}_k-\frac{\pi}{K_{\omega}}, \hat{\omega}_k+\frac{\pi}{K_{\omega}} \right]$ to find the estimate of $\bm{\psi}_k$ as
\begin{align}
    \hat{\bm{\psi}}_k = \arg \min _{\bm{\psi}_k}-\ln{L}(\bm{\psi}_k| \{\bm{\psi}_j\}_{j=1, j\neq k}^{K}).
\end{align}

\subsection{Model Order Determination}
For the case that the number of signals, i.e., the true model order $K$, is unknown, we consider extending the Bayesian information criterion (BIC) to the mixed-ADC data model to estimate the number of signals, and we refer to the approach as mBIC. Suppose that $\hat{\bm{\chi}}$ is the parameter vector estimated by the algorithm.

It is proven in Appendix \ref{Appendix:D} that the mBIC cost function for DOA and other unknown parameters has the following form:
\begin{align}\label{eq:85}
\text{mBIC}(\hat{K})= &-2 \ln{L}(\hat{\bm{\chi}}) 
+(2 \hat{K} N+3\hat{K}) \ln{MN} ,
\end{align}
where the estimate $\hat{K}$ of $K$ is selected as the integer that minimizes the above mBIC cost function with respect to the assumed number of targets $\hat{K}$. When Algorithm \ref{alg:1} is combined with mBIC we obtain Algorithm \ref{alg:2}.

\textcolor{black}{
\subsection{Complexity Analysis}
Regarding the SLIM algorithm, constructing $\hat{\mathbf{R}}_{t}^{i+1}$ and calculating $(\hat{\mathbf{R}}_{t}^{i+1})^{-1}$ entail complexities of $\mathcal{O}(M^2K_{\omega})$ and $\mathcal{O}(M^3)$, respectively. The update of $\hat{\zeta}_{t+1}^{i+1}$ requires computing $u$ and $v$, with a complexity of $\mathcal{O}(M^2N+MN)$. For updating $\mathbf{B}_{t+1}^{i+1}$, the complexity is $\mathcal{O}(M^2N+K_{\omega}MN)$. Therefore, the overall computational complexity of SLIM is $\mathcal{O}(I_1(M^2K_{\omega}+M^3+M^2N+K_{\omega}MN))$, where $I_1$ denotes the iteration number. The subsequent SLIM\&RELAX algorithm can be executed using an interior-point based bounded optimization method (e.g., "fmincon" of MATLAB), where each RELAX-based fine search's computational cost is proportional to $\mathcal{O}((2N+1)^{3.5})$. Assuming $I_2$ iterations are required for the RELAX-based refined progress, the total computational complexity of SLIM\&RELAX for a given target number is $\mathcal{O}(I_1(M^2K_{\omega}+M^3+M^2N+K_{\omega}MN)+I_2(2N+1)^{3.5})$. It is worth mention that, by using the technology of Gohberg-Semencul factorization and fast Fourier transform \cite{xue2011iaa,zhang2012fast}, the complexity of the SLIM algorithm can be further reduced to $\mathcal{O}(I_1(M^2+NK_{\omega}\log{K_{\omega}}))$.}

\section{Numerical examples}\label{sec:7}
In this section, we present several numerical examples to demonstrate the effectiveness of the  CRB analysis, the placement optimization of the mixed-ADC architecture, and the proposed parameter estimation algorithms. The inter-element spacing of the ULA is  $d = \frac{1}{2}\lambda$, with $M =64$, and $M_0 = 10$ antennas are equipped with high-precision ADCs and $M_1 = 54$ antennas are equipped with one-bit ADCs. All examples were run on a PC with Intel(R) Core(TM) i7-6700 CPU @ 3.40GHz and 48.0 GB RAM.

We consider an example of three incident signals with $\theta_1 = 10^{\circ}, \theta_2 = 20^{\circ},\theta_3 = 25^{\circ}, p_1=1, p_2 =0.8,p_3=0.8$. For the $k$th target, the real and imaginary parts of the complex amplitudes of the temporel waveforms of incident signals are assumed to be $\sqrt{p_k/2}$. The SNR is defined as 
\begin{align}
    \text{SNR}(\mathrm{dB})=10 \log _{10}\left(\sum_{n=1}^N|s(n)|^2 /\left(N \sigma^2\right)\right).
\end{align}

\subsection{CRB Analysis for Mixed-ADC Architecture}
Here we investigate  how the placement of high-precision and one-bit ADCs affects the DOA estimation performance. We consider the following three sceneries, shown in  Fig. \ref{fig:arr}, for the mixed-ADC architecture:

\begin{enumerate}
    \item $\{\delta_i = 1\}_{i=1}^{10}$ and $\{\delta_i = 0\}_{i=11}^{64}$, where the first 10 ADCs are the high-precision ADCs, see Fig. 1(a);
    \item $\{\delta_i = 0\}_{i=1}^{27}$, $\{\delta_i = 1\}_{i=28}^{37}$ and $\{\delta_i = 0\}_{i=38}^{64}$, where the middle 10 ADCs correspond to the high-precision ADCs, see Fig. 1(b);
    \item $\{\delta_i = 1\}_{i=1}^{5}$, $\{\delta_i = 0\}_{i=6}^{59}$ and $\{\delta_i = 1\}_{i=60}^{64}$, which turn out to be the optimal ADC placement for the mixed-ADC architecture, with the  high-precision ADCs  distributed evenly around the two edges of the ULA, see Fig. 1(c).
\end{enumerate}  
The sceneries 1), 2), and 3) above are referred to as ``Mixed-ADC1", ``Mixed-ADC2", and ``Mixed-ADC3", respectively.

For the one-bit ADC system and mixed-ADC architecture, the antenna-varying threshold has the real and imaginary parts selected randomly and equally likely from a predefined eight-element set $\{-h_{\text{max}},-h_{\text{max}}+\Delta,\dots,h_{\text{max}}-\Delta, h_{\text{max}}\}$ with $h_{\text{max}}=\sqrt{p_{\text{o}}}$ and $\Delta=\frac{h_{\text{max}}}{7}$, where $\sqrt{p_{\text{o}}}$ is the average received signal power at I/Q channels and it can be easily measured by automatic gain control (AGC) circuit at the antenna output before quantization. \textcolor{black}{In our experiment, the asymptotic CRB is calculated by (\ref{eq:47}), and the exact CRB is calculated by the inversion of the FIM, i.e., (\ref{eq:24}).}

\begin{figure}[htbp]
\centering
\includegraphics[scale=0.6]{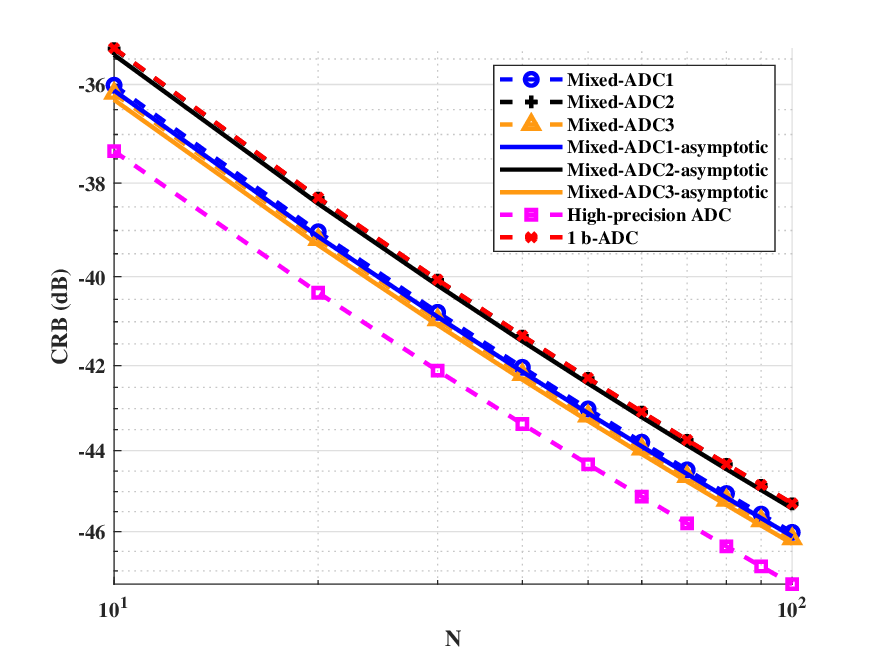}
\caption{CRB versus $N$, \textcolor{black}{SNR = -20dB}.}
\label{fig:1}
\end{figure}

Fig. \ref{fig:1} shows the CRBs versus $N$ when $\text{SNR} = -20$ dB. Note that the true CRB and the asymptotic CRB are very close to each other, and they both decrease with the number of snapshots. Also, ``Mixed-ADC3" in Fig. 1(c) gives the lowest CRB compared to the others, which agrees with our analysis.

\begin{figure}
    \centering
    \includegraphics[scale=0.6]{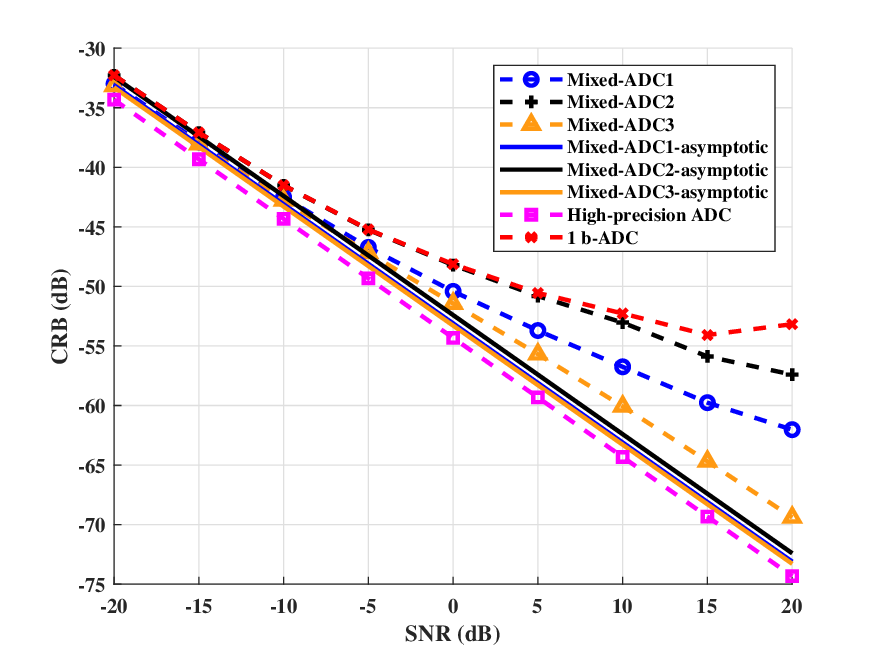}
    \caption{CRB of $\omega_1$ versus SNR, $N=5$.}
    \label{fig:2}
\end{figure}

Fig. \ref{fig:2} shows the CRBs versus SNR for $\omega_1$ \textcolor{black}{when the number of snapshot $N=5$}. Note that the CRB for the one-bit quantization is large in the high SNR region. Compared with the all one-bit ADC system, the mixed-ADC architectures  can achieve significant performance improvements for both asymptotic CRB and true CRB. The CRB of ``Mixed-ADC3" is nearly  $13$ dB lower than that of ``Mixed-ADC2".  The CRB of ``mixed-ADC2'' is very close to the CRB of a one-bit ADC system. This highlights the importance of proper high-precision and one-bit ADCs placements. The outstanding performance of ``Mixed-ADC3" can be attributed to having the largest aperture for high-precision ADCs. \textcolor{black}{Although the optimization goal is asymptotic CRB, the exact CRB performs better, especially in high SNR.}

\subsection{Ratio of the high-precision ADCs}

\begin{figure}[htbp]
    \centering
    \includegraphics[scale=0.45]{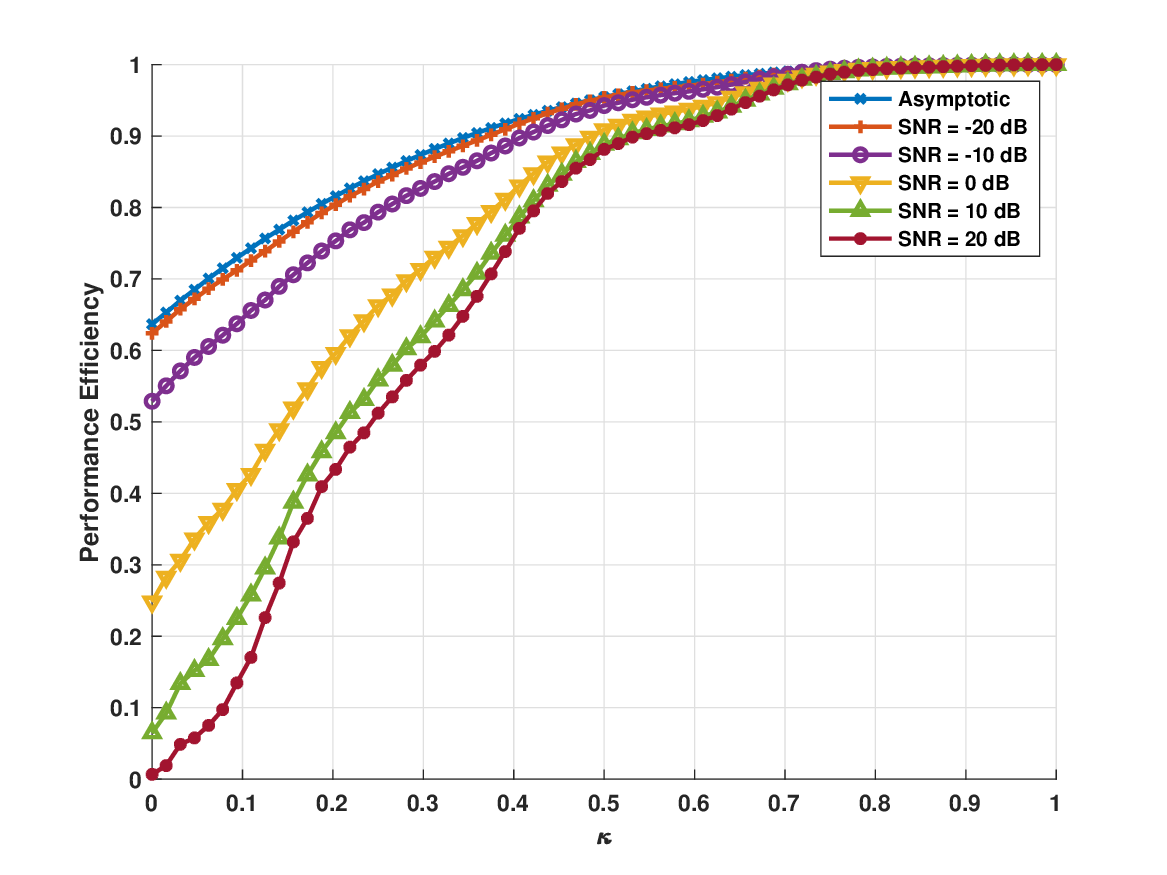}
    \caption{Performance efficiency versus the ratio of high-precision ADCs when $M=64, N=5$.}
    \label{fig:5}
\end{figure}

Fig. \ref{fig:5} plots the performance efficiency of the mixed-ADC architecture versus the ratio of the high-precision ADCs, where the total number of ADCs is $M=64$, SNR ranges from -20 dB to 20 dB, \textcolor{black}{and the number of snapshot $N=5$}. It is observed that, as the SNR increases, the performance efficiency of the architecture decreases. When $\kappa=0$, the all one-bit ADC system exhibits a decreasing trend from $2/\pi$ to 0. Notably, the performance efficiency experiences a sharp decline for small values of $\kappa$ in high SNR regimes, suggesting that increasing the number of high-precision ADCs can significantly enhance the performance of the mixed-ADC architecture. By selecting an appropriate ratio of high-precision ADCs, a balance can be achieved between cost and performance. In particular, we observe that achieving over 85\% capability of the high-precision system requires half of the total ADCs to be high-precision for high SNRs.

\subsection{Estimation Performance}

In our simulations, we set the number of grid point over frequency domain as $K_{w} = 10M$  \textcolor{black}{and the number of snapshot as $N=5$}. The practical convergence of SLIM is considered achieved when the relative change of power estimate $\mathbf{p}$ between two consecutive iterations is below a small threshold. For the outer MM iterations, we use the threshold $\varepsilon_1=10^{-6}$ and we terminate the iterations when a maximum iteration number $I_1 = 50$ is reached. For the cyclic iteration, we use the threshold $\varepsilon_2=10^{-4}$ and we terminate the iterations when the maximum iteration number $I_2 = 50$ is reached. For the RELAX based refined algorithm, we stop the iterations when the relative change of the negative log-likelihood function between two consecutive iterations is below $\varepsilon_3=10^{-6}$.

\begin{figure}[htbp]
    \centering
    \includegraphics[scale=0.6]{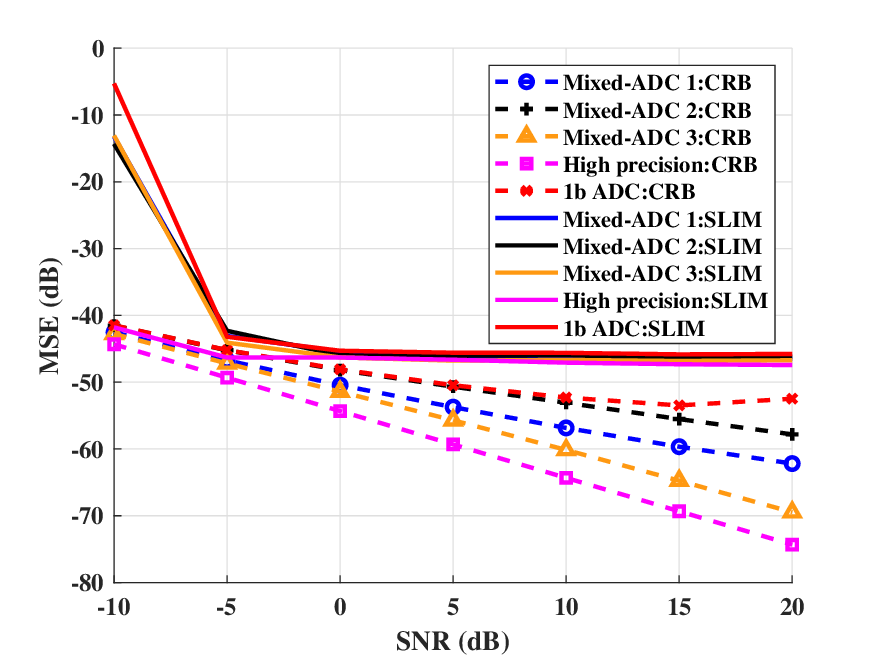}
    \caption{Comparison between CRBs and MSEs of the SLIM estimates of $w_1$ \textcolor{black}{in three different mixed-ADC placements, and systems with all one-bit and all high-precision ADCs.}}
    \label{fig:6}
\end{figure}

\begin{figure}[htbp]
    \centering
    \includegraphics[scale=0.6]{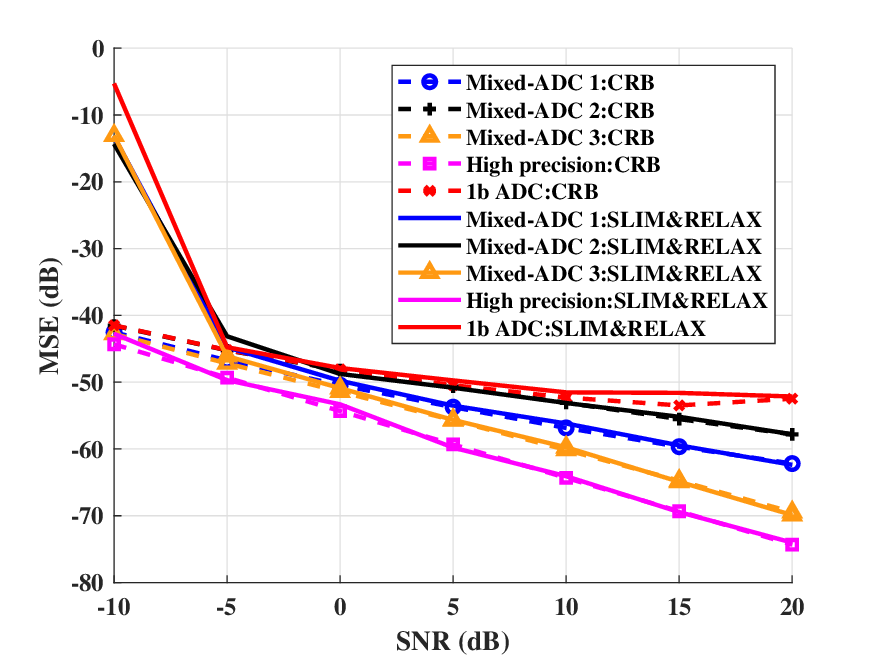}
    \caption{Comparison between CRBs and MSEs of the SLIM-RELAX estimates of $w_1$ \textcolor{black}{in three different mixed-ADC placements, and systems with all one-bit and all high-precision ADCs.}}
    \label{fig:7}
\end{figure}
\subsubsection{Example 1}
Fig. \ref{fig:6} and Fig. \ref{fig:7} compare the  mean-squared errors (MSEs) of $w_1$ obtained from two algorithms and the CRB with 100 Monte-Carlo trials for various placements of the ADCs, \textcolor{black}{and systems with all high-precision and all one-bit ADCs.}.  It is observed that the MSEs of SLIM reach a plateau when the SNR is above 5 dB. This behavior is due to the bias caused by the grid, which limits further reduction of the MSEs. However, the SLIM-RELAX algorithm demonstrates a significant improvement over SLIM, as the MSEs are able to approach the CRB as the SNR increases. \textcolor{black}{At the same time, all high-precision and all one-bit systems provide the upper and lower benchmark for the mixed-ADC architecture.}

\subsubsection{Example 2}
\begin{figure}[htbp]
    \centering
    \includegraphics[scale=0.45]{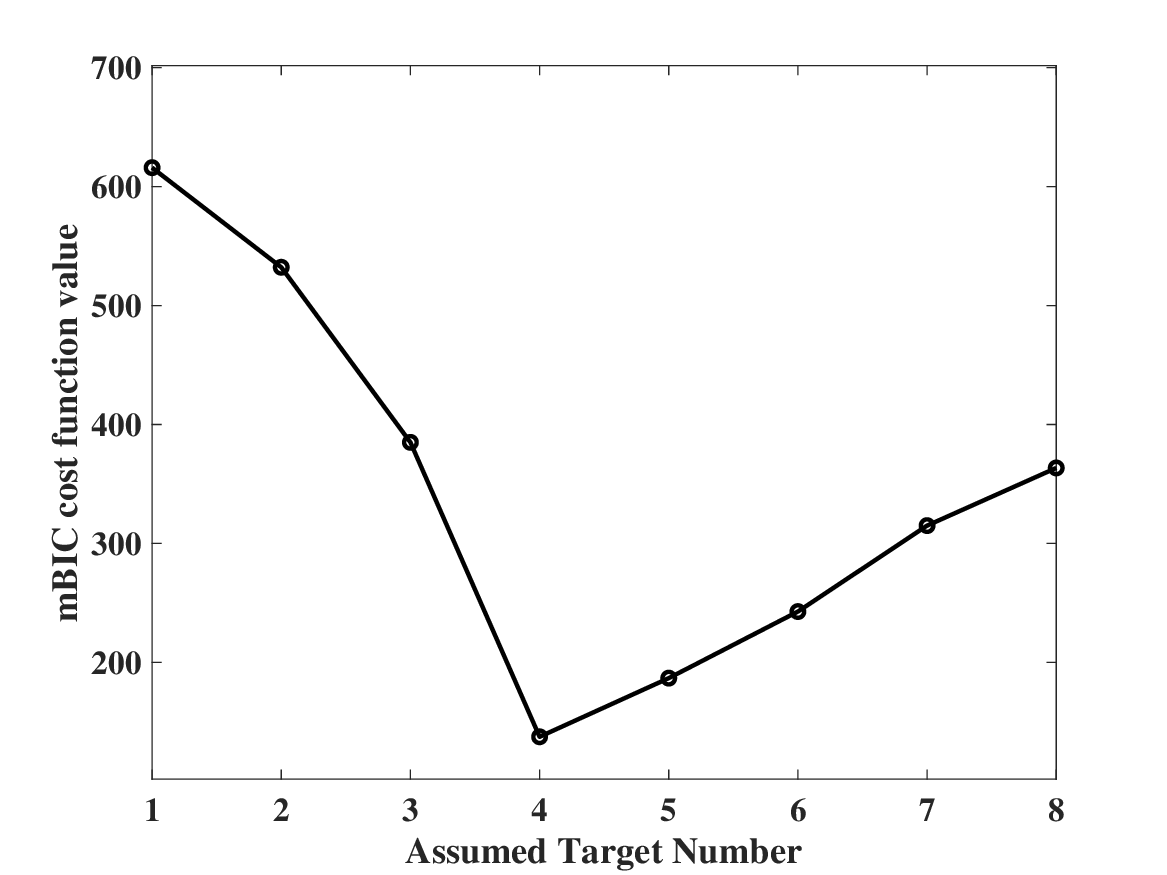}
    \caption{mBIC cost function value as a function of assumed target number when $M=64$, $M_0$ = 10 and SNR = 10 dB.}
    \label{fig:13}
\end{figure}

\begin{figure}[htbp]
    \centering
    \includegraphics[scale=0.45]{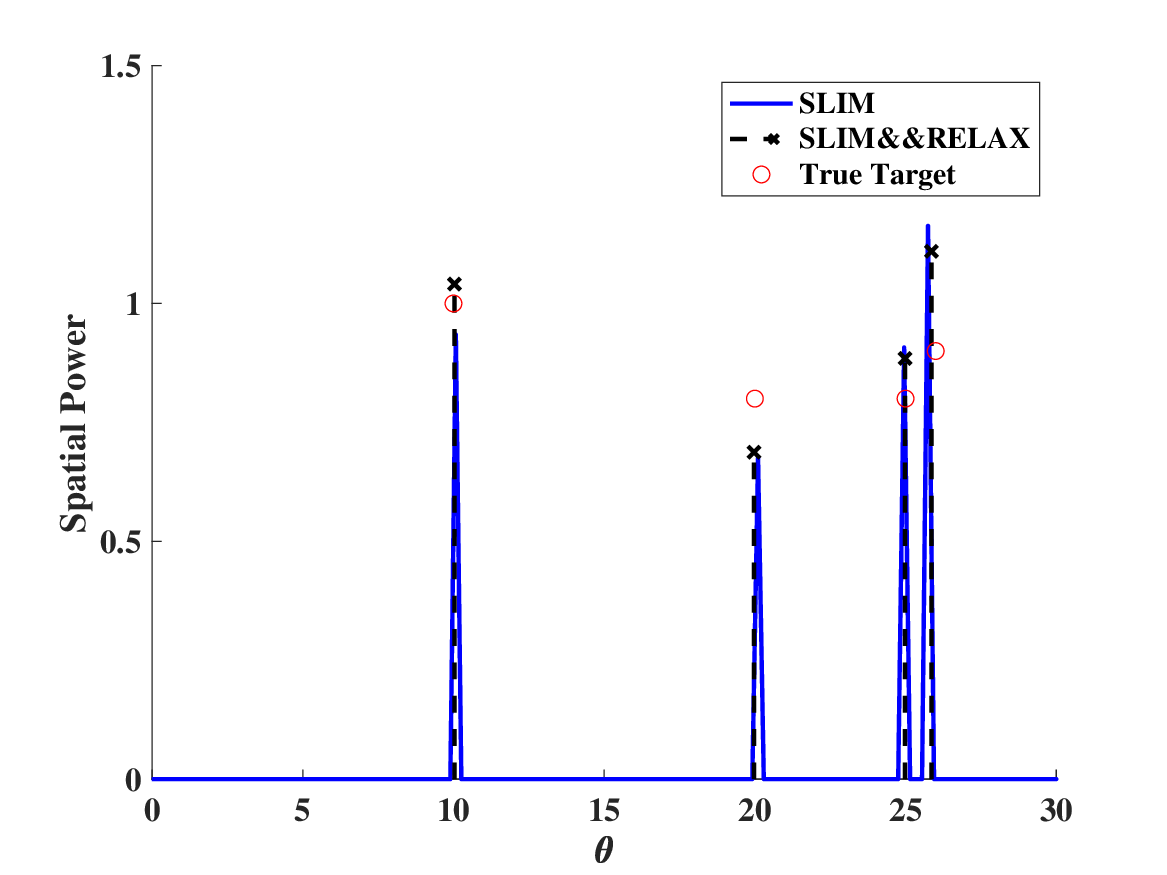}
    \caption{The spacial spectrum of the SLIM estimates (marked as blue) and the reﬁned results from SLIM-RELAX (marked as black $\times$), as compared with the true values (marked as red $\bigcirc$), when  $M=64$, $M_0=10$ and SNR = 10 dB.}
    \label{fig:12}
\end{figure}

Consider now the case of four targets with DOAs $\theta_1 = 10^{\circ}, \theta_2 = 20^{\circ},\theta_3 = 25^{\circ}, \theta_4 = 26^{\circ}$ and powers $p_1=1, p_2 =0.8,p_3=0.8, p_4=0.9$. It can be observed from Fig. \ref{fig:13} that the target number is correctly determined to be 4 when using SLIM-RELAX with mBIC under the mixed-ADC architecture with the optimal ADC placement. Moreover, Fig. \ref{fig:12} shows that the refined spatial spectrum obtained by SLIM-RELAX is more accurate, compared to that obtained by SLIM.

\section{Conclusions}
We have considered the mixed-ADC  architecture for DOA estimation using the ULA. We have derived the CRB and its lower bound. Then the asymptotic CRB is derived based on the lower bound of the CRB. We found that the asymptotic CRB is affected by the placement of high-precision and one-bit ADCs, and proved that the high-precision ADCs should be distributed evenly around the edges of the ULA to lower the CRB. This result can be extended to more general cases in which the ULAs are equipped with any two types of ADCs with different quantization precisions. To achieve the CRB, we have introduced a two-step DOA estimation algorithm to realize the ML estimation. Firstly, the sparse algorithm SLIM is extended to obtain accurate estimates under the mixed-ADC architecture. To further enhance the accuracy, we have used RELAX to cyclically refine the SLIM results. Numerical examples have been presented to demonstrate the effectiveness of the asymptotic CRB analysis, the optimal ADC placement, and the proposed DOA estimation algorithm for mixed-ADC architecture.

% \newpage
\appendices
\section{Proof of asymptotic CRB}
\label{Appendix:A}
From \cite{stoica1989maximum}, we have the following result:
\begin{align} \label{eq:86}
        \lim_{M\to\infty}\frac{1}{M^{r+1}}\sum_{t=1}^Mt^re^{\mathrm{i}tw}  =  0  \qquad w \in (0, 2\pi).
\end{align}
% Using (\ref{eq:43}) and the fact $M_0$ is bounded, we have:
Note that 
\small
\begin{align}
    \sum_{t=1}^M\frac{t^r}{M^{r+1}}e^{\mathrm{i}tw}g_t =  \frac{2}{\pi}\sum_{t=1}^M\frac{t^r}{M^{r+1}}e^{\mathrm{i}tw}+c\sum_{t\in \mathcal{M}_0}\frac{t^r}{M^{r+1}}e^{\mathrm{i}tw} ,
\end{align}
\normalsize
where $g_t \in \{1, \frac{2}{\pi}\}$, $c = 1-\frac{2}{\pi}$ and $\mathcal{M}_0$ denotes the set of indices of the high-precision ADCs. Using (\ref{eq:86}) and the fact that $M_0$ is constrained, we have 
\begin{align}
    \lim_{M\to\infty}\sum_{t=1}^M\frac{t^r}{M^{r+1}}e^{\mathrm{i}tw}g_t = 0 \qquad w \in (0, 2\pi).
\end{align}

Furthermore, for $k\neq p$, we have
\begin{align}
    &\frac{1}{M^3}\left[\dot{\mathbf{A}}^H\mathbf{\Sigma}_0\dot{\mathbf{A}}\right]_{kp} = \frac{a_{kp}}{M^3}\sum_{t=0}^{M-1}t^2e^{\mathrm{i}t(\omega_p-\omega_k)}g_t \underset{M\to \infty}{\longrightarrow} 0 ,\nonumber\\
    &\frac{1}{M^2}\left[\dot{\mathbf{A}}^H\mathbf{\Sigma}_0\mathbf{A}\right]_{kp} = \frac{b_{kp}}{M^2}\sum_{t=0}^{M-1}te^{\mathrm{i}t(\omega_p-\omega_k)}g_t \underset{M\to \infty}{\longrightarrow} 0 ,\nonumber\\
    &\frac{1}{M}\left[\mathbf{A}^H\mathbf{\Sigma}_0\mathbf{A}\right]_{kp} = \frac{1}{M}\sum_{t=1}^Me^{\mathrm{i}t(\omega_p-\omega_k)}g_t \underset{M\to \infty}{\longrightarrow} 0,
\end{align}
where  $a_{kp} = (k-1)(p-1)$ and $b_{kp} = -\mathrm{i}(k-1)$.

Hence $\frac{1}{M^3}\dot{\mathbf{A}}^H \mathbf{\Omega} \dot{\mathbf{A}}$ can be expressed as a diagonal matrix for a sufficiently large $M$. Note that the expression of $S$ is a quartic function of $M$, which completes the proof.
\section{Proof of optimal ADC arrangement}
\label{Appendix:B}
Let
\begin{align} \label{eq:46}
    H(m, n)  \triangleq \sum_{j=1, j\neq m,n}^Mg_j(2j-m-n).
\end{align}
Note that  (\ref{eq:46}) does not change if $g_m = g_n = 1$, and the above equation can be expressed as:
\small
\begin{align}
    &H(m, n) = 2\sum_{j=1}^Mg_j j - (m+n)\sum_{j=1}^Mg_j \nonumber\\
    & \geq 2\sum_{j=1}^mj +\frac{4}{\pi}\sum_{j=m+1}^{n-1}j + 2\sum_{j=n}^{n+M_0-m}j \nonumber\\
    &+\frac{4}{\pi}\sum_{j=n+M_0-m+1}^Mj -(m+n)\sum_{j=1}^Mg_j  \triangleq H^{\prime}(m,n).
\end{align}
\normalsize
In the first situation, using the fact that $ 0<m \leq M_\mathrm{h}$ and $m<n \leq M_\mathrm{h}+M_1$, we get 
\small
\begin{align}
    &H^{\prime}(m,n) - H^{\prime}(m+1,n) \nonumber\\
    & =\left(1-\frac{2}{\pi}\right)(2n+3M_0-4m-1)+\frac{2}{\pi}M \nonumber \\ &\geq \left(1-\frac{2}{\pi}\right)(3M_0-2M_\mathrm{h}+1)+\frac{2}{\pi}M > 0,
\end{align}
\normalsize
and
\small
\begin{align}
    &H^{\prime}(m,n) - H^{\prime}(m,n+1) \nonumber\\
    &= \left(1-\frac{2}{\pi}\right)(2m-M_0+1)+\frac{2}{\pi}M \nonumber \\
    &> \frac{2}{\pi}M-\left(1-\frac{2}{\pi}\right)M_1> \left(1-\frac{2}{\pi}\right)M_0 >0.
\end{align}
\normalsize
By combining the above inequalities,  we obtain
\small
\begin{align} \label{eq:67}
    H(m,n)& \geq H^{\prime}(m,n) \geq H^{\prime}(M_\mathrm{h}, M_\mathrm{h}+M_1) \nonumber \\
    & = M(M_0-2M_\mathrm{h}+1)+\frac{2}{\pi}(2M_\mathrm{h}+M_1) \nonumber \\
    & = \left\{\begin{array}{rcl}
        \left(1+\frac{2}{\pi}\right)M & & M_0 \quad \text{is even}\\
        \frac{2}{\pi}(M+1) & & M_0 \quad \text{is odd}
        \end{array} \right. > 0.
\end{align}
\normalsize
Therefore, we have $\tilde{S} > S$ for the first situation.

In the second situation, by letting $j^{\prime} = M+1-j$, we get
\small
\begin{align}
    H(m, n) &= 2\sum_{j=1}^Mg_{j}(M+1-j')-(2M+2-m^{\prime}-n^{\prime})\sum_{j=1}^Mg_j \nonumber\\
    & = (m^{\prime}+n^{\prime})\sum_{j=1}^Mg_j-2\sum_{j=1}^Mg_{j}j^{\prime}.
\end{align}
\normalsize
Due to the fact that $m^{\prime} \leq M_\mathrm{h}$, it is the same as the first case. Thus, we conclude that $H(m,n)<0$ for the second case.

\textcolor{black}{\section{Construction of majorizing function for $g(\mathbf{B}, \zeta)$}
\label{Appendix:C}
We rewrite function $g(\mathbf{B}, \zeta)$ as:
\small
\begin{align}
     g(\mathbf{B}, \zeta) = l_1(\mathbf{B}, \zeta) + l_0(\mathbf{B}, \zeta) + \sum_{r=1}^{K_{\omega}}\frac{2}{q}(\|\mathbf{b}_r\|_2^q-1),  
\end{align}
\normalsize
where $l_1(\mathbf{B}, \zeta) = -\ln{L}_1(\mathbf{B}, \zeta)$ and $l_0(\mathbf{B}, \zeta) = \frac{1}{2}\|\zeta \mathbf{Y_0}-\mathbf{A}_0\mathbf{B}\|_{\rm F}^2-M_0N\ln{\zeta^2}$. The construction of the majorizing function for $g(\mathbf{B}, \zeta)$ is based on two components:
\subsubsection{Majorizing Function for \texorpdfstring{$l_1(\mathbf{B}, \zeta)$}{Lg}}
For notational simplicity, let
\begin{align}\label{eq:64}
    \gamma_{\rm R}(m, n) &= Y_{1,\rm R}(m,n) \left(\Re\{\mathbf{a}_{1,m}^T\mathbf{b}(n)\}-\zeta H_{1,\rm R}(m,n)\right), \nonumber\\
    \gamma_{\rm I}(m, n) &= Y_{1,\rm I}(m,n) \left(\Im\{\mathbf{a}_{1,m}^T\mathbf{b}(n)\}-\zeta H_{1,\rm I}(m,n)\right), \nonumber\\
    f(x) & = -\ln{\Phi(x)}.
\end{align}
By utilizing the second-order Taylor expansion of $f(x)$ at $\hat{x}$ and the fact that $f^{''}(x) < 1$ for any real-valued $x$,  we have:
\begin{align}
    f(x) & \leq f(\hat{x}) + f^{'}(\hat{x})(x-\hat{x}) + \frac{1}{2}(x-\hat{x})^2 
    \nonumber \\
    & = \frac{1}{2}[x-(\hat{x}-f^{'}(\hat{x}))]^2 + \text{const}.
\end{align}
\\
By ignoring the constant term, the majorization function for $l_1(\mathbf{B}, \zeta)$ can be constructed as:
\begin{align}\label{eq:65}
    l_1(\mathbf{B}, \zeta) &= \sum_{n=1}^N\sum_{m=1}^{M_1}f(\gamma_{\rm R}(m, n))+\sum_{n=1}^N\sum_{m=1}^{M_1}f(\gamma_{\rm I}(m, n)) \nonumber \\
    & \leq \frac{1}{2}\|\bm{\gamma} - (\bm{\hat{\gamma}}^{i}- f^{'}(\bm{\hat{\gamma}}^{i})) \|_F^2,
\end{align}
where $\bm{\hat{\gamma}}^{i}$ is the estimate of $\bm{\gamma}$ at the $i$th outer MM iteration.
Along with (\ref{eq:64}), the majorizing function $G_1(\mathbf{B},\zeta|\hat{\mathbf{B}}^i, \hat{\zeta}^i)$ for $l_1(\mathbf{B}, \zeta)$ can be rewritten in a more compact form as
\begin{align}
    G_1(\mathbf{B},\zeta|\hat{\mathbf{B}}^i, \hat{\zeta}^i) = \frac{1}{2}\|\mathbf{A}_1\mathbf{B} - (\zeta \mathbf{H}+\hat{\mathbf{D}}^i)\|_{\rm F}^2,
\end{align}
where $\hat{\mathbf{D}}^i \in \mathbb{C}^{M_1\times N}$ with 
\small
\begin{align} \label{eq:72}
    &\hat{\mathbf{D}}^i(m, n) = Y_{1,\rm R}(m,n)\xi_{\rm R}^i(m,n)+\mathrm{i}Y_{1,\rm I}(m,n)\xi_{\rm I}^i(m,n), \nonumber \\
    &\xi_{\rm R/I}^i(m,n) = \hat{\gamma}_{\rm R/I}^i(m,n) - f^{'}(\hat{\gamma}_{\rm R/I}^i(m,n)), \nonumber \\
    &\hat{\gamma}^i_{\rm R}(m,n) = Y_{1,\rm R}(m,n) \left(\Re\{\mathbf{a}_{1,m}^T\hat{\mathbf{b}}^i(n)\}-\hat{\zeta}^i H_{1,\rm R}(m,n)\right), \nonumber \\
    &\hat{\gamma}^i_{\rm I}(m,n) = Y_{1,\rm R}(m,n) \left(\Im\{\mathbf{a}_{1,m}^T\hat{\mathbf{b}}^i(n)\}-\hat{\zeta}^i H_{1,\rm I}(m,n)\right).
\end{align}
\normalsize
\\
\subsubsection{Majorizing Function for \texorpdfstring{$\sum_{r=1}^{K_{\omega}}\frac{2}{q}(\|\mathbf{b}_r\|_2^q-1)$}{Lg}}
Note that $\sum_{r=1}^{K_{\omega}}\frac{2}{q}(\|\mathbf{b}_r\|_2^q-1)$ is a concave function of $\mathbf{B}$. At the $(i+1)$th MM iteration, the majorizing function for it can be constructed by the first-order Taylor expansion at $\hat{\mathbf{B}}$:
\begin{align}
    \sum_{r=1}^{K_{\omega}}\frac{2}{q}(\|\mathbf{b}_r\|_2^q-1) \leq & \sum_{r=1}^{K_{\omega}}\frac{\|\mathbf{b}_r\|_2^2}{\|\hat{\mathbf{b}}_{r}^{i}\|_2^{2-q}}+\text{const} \nonumber \\
     = & \frac{1}{N}\Tr(\mathbf{B}^H(\hat{\mathbf{P}}^i)^{-1}\mathbf{B})+\text{const},
\end{align}
where $\hat{\mathbf{P}}^i=\text{diag}\left\{\hat{p}_{1}^i, \hat{p}_{2}^i, \cdots, \hat{p}_{ K_{\omega}}^i\right\}$ with
\begin{align}\label{eq:74}
    \hat{p}_{r}^i=\frac{1}{N}\|\hat{\mathbf{b}}_{r}^{i}\|_2^{2-q} .
\end{align}
\\~~ Invoking the above majorizing function and ignoring the irrelevant term, we can obtain the following majorizing function for $g(\mathbf{B}, \zeta)$ at the $(i+1)$th iteration:
\small
\begin{align}\label{eq:75}
    &G(\mathbf{B},\zeta|\hat{\mathbf{B}}^i,\hat{\zeta}^i) =  \frac{1}{2}\|\mathbf{A}_1\mathbf{B} - (\zeta \mathbf{H}+\hat{\mathbf{D}}^i)\|_{\rm F}^2 \nonumber \\&+ \frac{1}{2}\|\zeta\mathbf{Y}_0-\mathbf{A}_0\mathbf{B}\|_{\rm F}^2+\frac{1}{N}\Tr(\mathbf{B}^H(\hat{\mathbf{P}}^i)^{-1}\mathbf{B})-2M_0N\ln{\zeta}.
\end{align}
\normalsize
}
\section{Derivation of mBIC}
\label{Appendix:D}
We rewrite mixed-ADC signal model as
\begin{equation}
   \mathbf{Y} = \mathbf{Z} \odot (\boldsymbol{\bar{\delta}} \otimes \vec{1}_N^T) + \mathbf{X} \odot (\boldsymbol{\delta} \otimes \vec{1}_N^T) \in \mathbb{C}^{M\times N}
\end{equation}
with the unknown parameter vector being $\bm{\chi}=\begin{bmatrix}\bm{\omega}^T,\ \vec{s}_{\rm R}^T, \ \vec{s}_{\rm I}^T, \ \sigma\end{bmatrix}^T\in \mathbb{R}^{(K+2KN+1)\times 1}$.

By assuming that the prior probability density function (PDF) of $\bm{\chi}$ is flat around the maximum likelihood estimate and independent of $M$ and $N$, the estimate of the signal model order $K$ can be derived by minimizing the following criterion \cite{1311138}
\begin{align}
    -2 \ln L(\mathbf{Y}, \hat{\boldsymbol{\chi}})+\ln |\hat{\mathbf{F}}(\boldsymbol{Y}, \hat{\boldsymbol{\chi}})|,
\end{align}
where $L(\mathbf{Y}, \hat{\bm{\chi}})$ is the log-likelihood function of $\mathbf{Y}$ and $\hat{\bm{\chi}}$ is the ML estimate of $\bm{\chi}$. The matrix $\hat{\mathbf{F}}(\mathbf{Y}, \hat{\bm{\chi}})$ is defined as 
\begin{align}
    \hat{\mathbf{F}}(\mathbf{Y}, \hat{\bm{\chi}})=-\left.\frac{\partial^2 \ln L(\mathbf{Y}, \bm{\chi})}{\partial \bm{\chi} \partial \bm{\chi}^T}\right|_{\bm{\chi}=\hat{\bm{\chi}}}.
\end{align}
Under mild conditions (see, e.g., \cite{1311138} and the references therein), the matrix $\hat{\mathbf{F}}(\mathbf{Y}, \hat{\bm{\chi}})$ has the following asymptotic relationship with the FIM $\mathbf{F}$ for the parameter vector $\bm{\chi}$ \cite{li2018bayesian}
\begin{align}
    \left[\mathbf{P}_N \mathbf{F} \mathbf{P}_N-\mathbf{P}_N \hat{\mathbf{F}}(\mathbf{Y}, \hat{\bm{\chi}}) \mathbf{P}_N\right] \rightarrow 0 \text { as } N \rightarrow \infty,
\end{align}
where $\mathbf{P}_N$ is a normalization matrix, and
\begin{align}
    \mathbf{F}=\mathbb{E}\left\{-\frac{\partial^2 \ln L(\mathbf{Y}, \bm{\chi})}{\partial \bm{\chi} \partial \bm{\chi}^T}\right\}.
\end{align}

Since the noise term has little influence on BIC, for computational simplicity, we primarily focus on the situation that the noise term is known, where the unknown parameter is $\bm{\varphi} = \begin{bmatrix}\bm{\omega}^T,\ \vec{s}_{\rm R}^T, \ \vec{s}_{\rm I}^T \end{bmatrix}^T\in \mathbb{R}^{(K+2KN)\times 1}$.

Let us revisit the FIM $\mathbf{F}_m$ for the mixed-ADC data in (\ref{eq:20}). Because the function $B(\cdot)$ defined in (\ref{eq:19}) is positive and bounded by $(0, 4]$ \cite{li2018bayesian}, we have the following inequality: 
\begin{align}
    \mathbf{F}_1 \preceq \mathbf{F}_m \preceq \mathbf{F}_0.
\end{align}

Note that the BIC cost functions for the high-precision and  one-bit systems are the same. Hence the BIC cost function for the mixed-ADC architecture system is given by
\begin{align}
    \text{mBIC}(K) = -2 \ln L(\mathbf{Y}, \hat{\boldsymbol{\chi}})+(2KN+3K)\ln{MN}.
\end{align}

% By letting
% \begin{align}
%     \tilde{\mathbf{F}} = \frac{2}{\sigma^2}\left(\tilde{\mathbf{U}}_{\rm R}\tilde{\mathbf{U}}_{\rm R}^T+\tilde{\mathbf{U}}_{\rm I}\tilde{\mathbf{U}}_{\rm I}^T\right),
% \end{align}
% we have the fact that $\ln{|\tilde{\mathbf{F}}|}$ and $\ln{|\mathbf{F}_m|}$ are asymptotically equivalent (see more detail in \cite{li2018bayesian}), to within a constant that does not depend on $N$. Therefore, the FIM $\mathbf{F}_m$ can be substituted by a simpler matrix $\tilde{\mathbf{F}}$. Note that the $\tilde{\mathbf{F}}$ is not equal to the FIM for the conventional BIC. Next, we analyze $\tilde{\mathbf{F}}$ and the normalization matrix $\mathbf{P}_N$ needed for $\mathbf{P}_N \mathbf{F} \mathbf{P}_N$ to be $O(1)$ for our mixed-ADC based architecture signal model.
% and
% \begin{align}
%     c_1 &= \min _{1 \leqslant k \leqslant MN}\left\{B\left(\frac{\Re\left(\zeta_{k}\right)}{\sigma / \sqrt{2}}\right),B\left(\frac{\Im\left(\zeta_{k}\right)}{\sigma / \sqrt{2}}\right)\right\}, 
%     % \nonumber \\
%     % c_2 &= \max _{1 \leqslant k \leqslant MN}\left\{B\left(\frac{\Re\left(\zeta_{k}\right)}{\sigma / \sqrt{2}}\right),B\left(\frac{\Im\left(\zeta_{k}\right)}{\sigma / \sqrt{2}}\right)\right\}.
% \end{align}
% We have the following matrix inequalities
% \begin{align}
%   \frac{c_1}{2\pi}\tilde{\mathbf{F}}  \preceq \mathbf{F}_m(\bm{\chi}) \preceq \frac{2}{\pi}\tilde{\mathbf{F}}
% \end{align}

\bstctlcite{IEEEexample:BSTcontrol}
\bibliographystyle{IEEEtran}
\small
\bibliography{ref}
\end{document}